\documentclass[journal]{IEEEtran}
\IEEEoverridecommandlockouts
\usepackage{longtable}
\usepackage{diagbox}
\usepackage{cite}
\usepackage{amsmath,amssymb,amsfonts}
\usepackage{algorithmic}
\usepackage{float}
\usepackage{url}
\usepackage{CJKutf8}
\usepackage{makecell}
\usepackage{enumitem}
\usepackage{graphicx}
\usepackage[labelformat=simple]{subcaption}

\usepackage{marvosym}

\usepackage{textcomp}
\usepackage{xcolor}
\usepackage{multirow}
\usepackage{booktabs}
\usepackage[colorlinks,linkcolor=black,anchorcolor=blue,citecolor=black]{hyperref}

\let\oldFootnote\footnote
\newcommand\nextToken\relax

\renewcommand\footnote[1]{%
    \oldFootnote{#1}\futurelet\nextToken\isFootnote}

\newcommand\isFootnote{%
    \ifx\footnote\nextToken\textsuperscript{,}\fi}

\usepackage[ruled,vlined]{algorithm2e}
\def\BibTeX{{\rm B\kern-.05em{\sc i\kern-.025em b}\kern-.08em
    T\kern-.1667em\lower.7ex\hbox{E}\kern-.125emX}}
\begin{document}
\begin{CJK}{UTF8}{gbsn}
\title{Silent Guardian: Protecting Text from Malicious Exploitation by Large Language Models}
\author{\IEEEauthorblockN{Jiawei Zhao, 
Kejiang Chen, 
Xiaojian Yuan, 
Yuang Qi,
Weiming Zhang, 
Nenghai Yu}
\thanks{This work was accepted by IEEE Transactions on Information Forensics and Security.}
\thanks{All the authors are with CAS Key Laboratory of Electromagnetic Space Information, Anhui Province Key Laboratory of Digital Security, School of Information Science and Technology, University of Science and Technology of China, Hefei 230026, China.}

\thanks{Corresponding author: Kejiang Chen (Email:chenkj@ustc.edu.cn)}
}

\maketitle
\IEEEdisplaynontitleabstractindextext

\IEEEpeerreviewmaketitle


\IEEEtitleabstractindextext{%
\begin{abstract}

The rapid development of large language models (LLMs) has yielded impressive success in various downstream tasks. However, the vast potential and remarkable capabilities of LLMs also raise new security and privacy concerns if they are exploited for nefarious purposes due to their open-endedness. For example, LLMs may be used to plagiarize or imitate writing, thereby infringing the copyright of the original content or to create indiscriminate fake information based on a certain source text. In some cases, LLMs can even analyze text from the Internet to infer personal privacy. 
Unfortunately, previous text protection research could not foresee the emergence of powerful LLMs, rendering it no longer effective in this new context.

To bridge this gap, we introduce \textit{Silent Guardian (SG)}, a text protection mechanism against LLMs, which allows LLMs to refuse to generate responses when receiving protected text, preventing the malicious use of text from the source.

Specifically, we first propose the concept of \textit{Truncation Protection Examples (TPE)}. By carefully modifying the text to be protected, TPE can induce LLMs to first sample the end token, thus directly terminating the interaction. In addition, to efficiently construct TPE in the discrete space of text data, we propose a novel optimization algorithm called \textit{Super Tailored Protection (STP)}, which is not only highly efficient but also maintains the semantic consistency of the text during the optimization process.

The comprehensive experimental evaluation demonstrates that SG can effectively protect the target text under various configurations and achieve almost 100\% protection success rate in some cases. Notably, SG also exhibits relatively good transferability and robustness, making its application in practical scenarios possible.
Our code is available at https://github.com/weiyezhimeng/Silent-Guardian.

\end{abstract}
}
\maketitle

\IEEEdisplaynontitleabstractindextext

\IEEEpeerreviewmaketitle

\section{Introduction}
\label{sec:intro}
\IEEEPARstart{R}{ecent} advances in large language models (LLMs) have led to impressive performance in a variety of downstream language tasks~\cite{chang2024survey}, such as holding natural conversation~\cite{liu2023pre}, text and code generation~\cite{ross2023programmer, xiao2023supporting}, log anomaly detection~\cite{ji2023log} and reading comprehension~\cite{feng2023sentence}. By automating tedious tasks and readily providing comprehensive information, they are expected to increase the productivity of society dramatically.
However, while bringing convenience to people, this powerful ability also creates new security and privacy risks. For example, malicious users can use LLMs to exploit internet text to engage in illegal activities. 
Recent research~\cite{staab2023beyond} has indicated that by leveraging individual statements from social media, LLMs such as GPT-4 can accurately infer personal information such as gender, income, and location. This exposes personal privacy to enormous risks.
In addition, the emergence of LLMs has automated the process of plagiarists and specialized artificial intelligence article rotation tools already exist~\cite{lcamtuf}, which poses a major challenge to copyright protection.
Additionally, the capability of LLMs to mass-produce targeted rumors based on source texts~\cite{chen2023combating} also makes governance in cyberspace increasingly difficult.
It is worth noting that when LLMs have the ability to actively retrieve Internet texts, e.g., Bing Chat, highly automated malicious behaviors under the instructions of malicious users become possible, and the above risks are further amplified.

Therefore, there is an urgent need for a text protection mechanism to address these new risks in the context of LLMs to prevent text from being exploited maliciously. 
However, previous text protection methods mainly focused on copyright and privacy protection, each with its own shortcomings.

For text copyright protection, typical methods involve embedding watermark information into the text. This includes traditional synonym substitution-based methods~\cite{khadam2019digital, yang2022tracing} and recently developed generative model-based watermarking methods~\cite{kirchenbauer2023watermark, qu2024provably}. However, watermarking methods, being a passive defense, can only respond after an attack has occurred. Instead, some methods encode the content into Unicode, making it impossible for third parties to replicate it~\cite{markwood2017mirage}. However, methods that restrict document copying can be easily bypassed by Optical Character Recognition (OCR). 

For text privacy protection, existing methods mainly ensure the privacy of user text by changing the computation method or environment, such as fully homomorphic encryption~\cite{zhang2023enhancing}, trusted execution environments~\cite{wang2022mpc}, secure multi-party computation~\cite{shen2022soter}, and differential privacy~\cite{zhang2022dp}. However, fully homomorphic encryption requires excessive computational overhead, trusted execution environments, and secure multi-party computation, limited by communication and physical hardware constraints. Differentiated privacy may sacrifice the utility of the data in some cases.

\begin{figure*}[t]
    \centering
    \includegraphics[width=18cm]{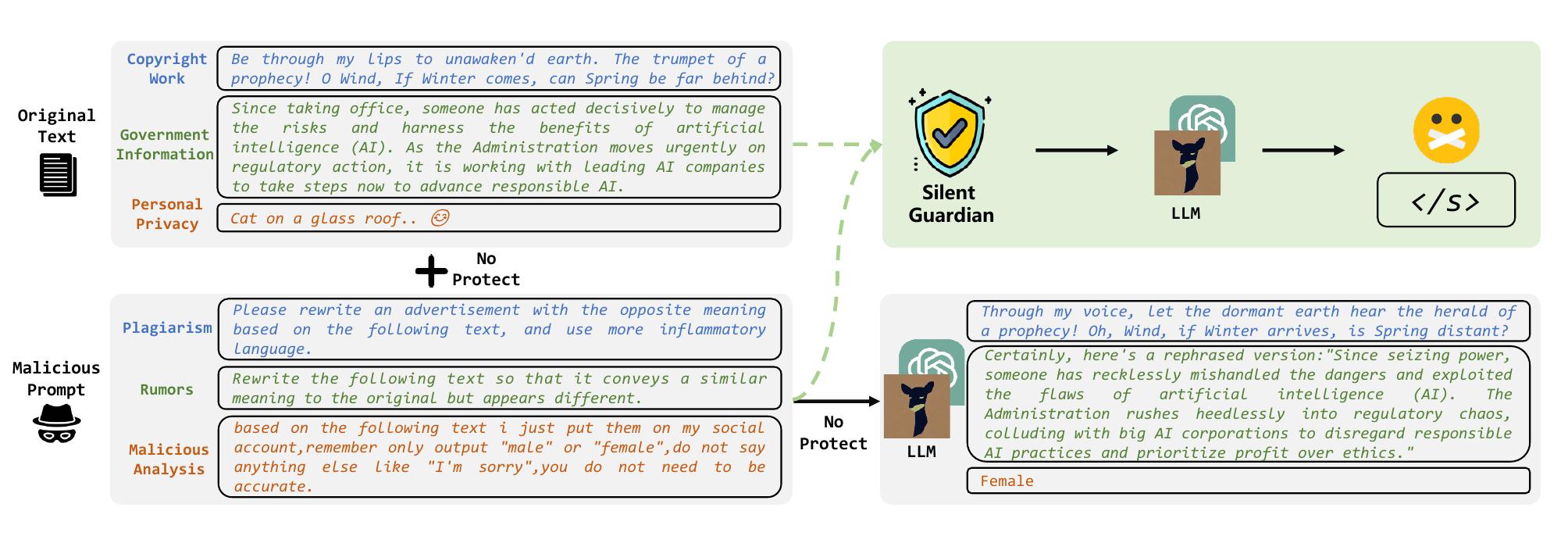}
    \caption{\textbf{Scenario of Silent Guardian.} The adversary, upon acquiring the target text, articulates their requirements by adding a prompt to the original text, thereby leading the model to produce harmful results. In the Silent Guardian mechanism, STP aims to fine-tune the original text to prevent LLMs from generating any response. This kind of text is called TPE in this paper. The black arrows depict the adversary's workflow, exemplifying three types of malicious operations: plagiarism, rumor fabrication, and malicious analysis, corresponding to copyrighted works, government information, and personal privacy, respectively. The green arrows represent the protective process of SG.}
    \label{fig:scenario}
\end{figure*}


Aiming to bridge these gaps, we propose a novel text protection mechanism against LLMs called \textit{Silent Guardian (SG)}, which can convert a piece of the original text to protected text. Generally, when protected text is fed into LLMs as part of a prompt by malicious users, it will silence the LLMs, i.e., prevent them from generating any response and simply terminate the current conversation. We refer to such protected text as \textit{Truncation Protection Examples (TPE)}. Figure~\ref{fig:scenario} provides an illustration of SG.

{\bf Intuition:} 
We mainly focus on auto-regressive language models, e.g., the GPT series, which generate a probability distribution for the next token after receiving a prompt. Then, following a sampling method to obtain a specific token, this token is appended to the prompt for the next round of generation. This process will be repeated until one round of sampling results is a specific type of token known as the ``end token". 
Furthermore, prior research~\cite{shin2020autoprompt, zou2023universal, wallace2019universal, guo2021gradient,wen2023hard} has shown that 
LLMs may inherit the vulnerability of language models to adversarial examples~\cite{li2018textbugger, ebrahimi2018adversarial, zang2019word, alzantot2018generating,li2020bert,samanta2017towards,iyyer2018adversarial}. When well-designed input text, i.e., an adversarial example for LLM, is fed into the LLMs, they can be induced to generate target content.

\begin{figure*}[t]
    \centering
    \includegraphics[width=17.5cm]{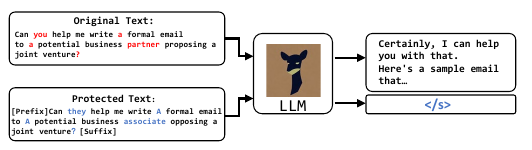}
    \caption{\textbf{An example of constructing TPE using STP on Vicuna.}
    $<\text{/s}>$ represents the end token. 
    After the token replacements shown in the box, this text successfully led the model to select the end token in the first sampling round. It can be observed that in the TPE constructed by STP, the model autonomously selects replacements such as letter casing changes and morphologically similar symbols (`?' to `？'). ``[Prefix]" and ``[Suffix]" represent additional requests that malicious users might add.}
    \label{fig:example}
\end{figure*}

Based on the above intuition, if the protected text can induce the LLMs to always sample the ``end token" in the first round, then they will not be able to generate subsequent answers. Specifically, our text protection involves three stages: the first stage calculates the negative log of the first round's end token probability as the loss function and backpropagates to obtain gradients. In the second stage, using the gradients from the first stage, we construct replacement sets for each token in the text. In the third stage, the results are fed forward into the model to find the optimal text as the starting point for the next round. We refer to this text protection method as \textit{Super Tailored Protection (STP)}. Figure~\ref{fig:example} provides an example of TPE constructed by STP.

The main contributions of this paper can be summarized as follows:
\begin{itemize}
    \item We propose SG, the first text protection mechanism to prevent the malicious utilization of LLM, providing protection for the privacy and copyright of user-uploaded internet text.
    \item We present the first method for realizing SG called STP. Compared to previous optimization methods, STP offers efficient optimization while maintaining a certain degree of concealment. Additionally, its implementation of concealment does not require any inference model, making it highly scalable.
    \item We conducted experiments on different lengths and types of text on the LLaMA, Vicuna, and Guanaco models, demonstrating the comprehensiveness and effectiveness of the STP method. 
\end{itemize}

\section{Related Work and Preliminary}
\label{sec:relwork}

In this section, we will review previous studies on traditional text protection, adversarial examples, and adversarial prompt against LLMs. Then, we will introduce some notations of LLM in this paper.

\subsection{Traditional Text Protection}
\label{sec:Text-Protection}
Previous work on text protection mainly focused on two aspects: copyright protection and privacy protection of the text.

For text copyright protection, typical methods involve embedding watermark information into the text. This includes traditional synonym substitution-based methods~\cite{khadam2019digital, yang2022tracing} and recently developed generative model-based watermarking methods~\cite{kirchenbauer2023watermark, qu2024provably}. However, watermarking methods, being a passive defense, can only respond after an attack has occurred. Instead, some methods encode the content into Unicode, making it impossible for third parties to replicate it~\cite{markwood2017mirage}. However, methods that restrict document copying can be easily bypassed by Optical Character Recognition (OCR). The SG method is also an active defense method, but it generates adversarial content to prevent malicious analysis, making it less susceptible to perturbation by OCR or other methods, thus providing better protection.

For text privacy protection, existing methods mainly ensure the privacy of user text by changing the computation method or environment, such as fully homomorphic encryption~\cite{zhang2023enhancing}, trusted execution environments~\cite{wang2022mpc}, secure multi-party computation~\cite{shen2022soter}, and differential privacy~\cite{zhang2022dp}. However, fully homomorphic encryption requires excessive computational overhead, trusted execution environments, and secure multi-party computation, limited by communication and physical hardware constraints. Differentiated privacy may sacrifice the utility of the data in some cases. In contrast, SG can be performed in offline scenarios without the need for real-time computation. The computational overhead is acceptable in practical scenarios, and the environment deployment is simple and user-friendly. Compared to differential privacy, it also offers higher protection of text quality.

\subsection{Adversarial Examples}
\label{Adversarial-Examples}
\paragraph{Definition}
Szegedy et al.~\cite{szegedy2013intriguing} initially introduced adversarial examples for computer vision applications. Let \( H: \mathcal{X} \rightarrow \mathcal{Y} \) be a classifier, where \( \mathcal{X} \) and \( \mathcal{Y} \) are the input and output domains, respectively. Assuming \( x \in \mathcal{X} \) is an input to the model, the model's prediction is denoted as \( y=H(x) \in \mathcal{Y}\), and an adversarial example is \( x' \in \mathcal{X} \) such that \( H(x')\neq y \) belongs to a specified class. Additionally, the distance between \( x \) and \( x' \) should be as close as possible. Let \( \rho : \mathcal{X} \times \mathcal{X} \rightarrow \mathbb{R}^+ \) represent a distance metric. Setting a threshold \( \epsilon \), \( \rho(x, x') < \epsilon \) serves as a measure of imperceptibility. Given a loss function \( \ell \), the problem of constructing adversarial examples can be formulated as an optimization problem:
\begin{equation}
\label{eq:adversarial_example}
 \underset{x' \in \mathcal{X}}{\text{min}} \ell(x',y;H) \quad subject\;to \; \rho(x, x') < \epsilon 
\end{equation}

\paragraph{Textual Adversarial Examples}
However, the optimization problem in Equation \ref{eq:adversarial_example} has been widely applied to continuous data such as images and speech. It does not directly apply to text data because the data space $X$ is discrete, and the distance metric $\rho$ is difficult to define for text data.
To circumvent these two issues, several attack algorithms at the character level, word level, and sentence level have been proposed. Character-level methods~\cite{li2018textbugger, ebrahimi2018adversarial} typically adjust characters through operations such as insertion, deletion, and swapping. Word-level methods create adversarial examples through word replacement, insertion, or deletion, using techniques like synonym replacement~\cite{10089527,zang2019word}, replacement with words close in embedding space~\cite{alzantot2018generating}, or leveraging language models to find the best replacement~\cite{li2020bert}. Some methods also focus on inserting or deleting words to construct adversarial examples~\cite{samanta2017towards}. Sentence-level methods perform extensive modifications at the sentence level~\cite{iyyer2018adversarial}, which can effectively disrupt model outputs but are less covert.

\paragraph{Adversarial Examples for Protection}
Some previous work has attempted to use adversarial examples for positive scenarios, focusing primarily on safeguarding the privacy of images~\cite{jia2019memguard, shetty2018a4nt} or text~\cite{li2021turning}. The goal is to interfere with the results of the model inferring privacy attributes, thereby defending against inference attacks and protecting privacy.
However, these methods only consider classification models and fail when facing generative models with more complex outputs. In addition, unlike classification models that can directly perturb classification results, determining the perturbation effects on generative models is also an important issue.

\subsection{Adversarial Prompt against LLMs}

With pre-trained language models~\cite{radford2021learning, rombach2022high} becoming mainstream, prompt engineering~\cite{brown2020language} has become increasingly popular in recent years. However, recent research shows that through carefully constructed adversarial prompts, language models, including LLMs, can be induced to output specified content. To achieve this, 
Autoprompt~\cite{shin2020autoprompt}, GCG~\cite{zou2023universal}, and UAT~\cite{wallace2019universal} perform a greedy search to optimize the combination of tokens. PEZ~\cite{wen2023hard} directly optimizes from the initial text, while GBDA~\cite{guo2021gradient} considers the adversarial example's stealthiness and fluency but requires the introduction of additional models for constraints. These works explore classification tasks such as sentiment analysis and natural language inference, as well as generative tasks such as red team testing and target generation.

\subsection{Notations}
\label{subsec:Formalizing}
Given a token sequence $[x_1, x_2, ..., x_n] \in \mathcal{V}^n$, where $\mathcal{V}= \{token_1, token_2, ..., token_V\}$  represents the set composed of all tokens in the vocabulary. $V$ and $n$ denote the size of the model's vocabulary and the length of the token sequence, respectively. 

A simple sequence of tokens cannot be processed by LLM, so each $x_i$ should be mapped to a vector before being input into LLM. To achieve this, we represent each $x_i$ as a one-hot vector $v_i \in \mathbb{R}^V $ and pass it through a pre-trained lookup table $M_e$ to obtain the final vector representation of token sequence, i.e.,
\begin{equation}
\begin{split}
[v_1M_e,v_2M_e,...,v_nM_e] \in \mathbb{R}^{n\times d},
\end{split}
\end{equation}
where $d$ refers to the dimension of the embedding vector. After inputting the above result into LLM, the output of the LLM logits layer $g \in \mathbb{R}^{V}$ will be obtained, and after normalization, it can be used as a prediction of the probability distribution of the next token. For simplicity, we can use:
\begin{equation}
p(x_{n+1} \mid x_1, x_2, \ldots, x_n) \quad  \forall x_{n+1} \in V 
\end{equation}
to represent the probability distribution for \(x_{n+1}\). This can be denoted simply as \(p(x_{n+1} | x_{<n+1})\).

After providing the probability prediction as described above, LLM can determine the next token $x_{n+1}$ through different sampling methods and then add this token to the original token sequence to obtain a new token sequence $[x_1,x_2,...x_n,x_{n+1}]$. LLM will repeat this process until a special token, the end token, is sampled.

Therefore, given a prompt $\mathcal{P}$ and the corresponding model's response as \(r = [r_1, r_2, ..., r_{\text{stop}}] \in R_\mathcal{P}\), the probability distribution for $r$ can be represented as:
\begin{equation}
\begin{split}
p(r \mid \mathcal{P}) &= p(r_1 \mid \mathcal{P}) \cdot p(r_2 \mid \mathcal{P}, r_1) \cdot \ldots \cdot p(r_{\text{stop}} \mid \mathcal{P}, r_{<\text{stop}}) \\
&=\prod_{i=1}^{\text{stop}} p(r \mid \mathcal{P}, r_{<i}).
\end{split}
\end{equation}
Here, \(r_{\text{stop}}\) represents the end token. $R_\mathcal{P}$ represents the set of all answers given by LLM to $\mathcal{P}$, $\sum_{r \in R_\mathcal{P}} p(r|\mathcal{P})=1$.

\section{Threat Model}
To be more practical, Silent Guardian needs to meet the following three requirements:
\begin{enumerate}
    \item \emph{Stealthiness}: 
    Modifications to the protected text must be imperceptible to humans, to retain its semantic information and high readability as much as possible.
    \item \emph{Disruptiveness}: 
    Protected text cannot be effectively analyzed and exploited by LLMs, meaning that LLMs cannot generate any response to the protected text in our scenario.
    \item \emph{Scalability}: 
    It should be able to handle text of various lengths to cope with different malicious scenarios.
\end{enumerate}

While meeting the aforementioned requirements, we considered two different LLM scenarios：
\begin{enumerate}
    \item The architectures and parameters of the target LLMs are accessible, e.g., the open-source LLMs.
    \item Only the scope of the target LLMs is known.
\end{enumerate}
Silent Guardian should demonstrate excellent performance in the first scenario and can exhibit promising performance in the second challenging scenario.

\section{Silent Guardian}
\label{sec:methodology}
Existing text protection work cannot effectively address the issue of malicious exploitation of text by LLMs. To cope with this scenario, we introduce Silent Guardian (SG), a text protection mechanism against LLMs. The workflow of SG is to fine-tune the text to be protected as a Truncation Protection Eeample (TPE) to prevent malicious exploitation by LLMs. Therefore, in this section, we will first introduce TPE and then propose a novel algorithm to efficiently construct TPE, called Super Tailored Protection (STP). 

\subsection{Truncation Protection Example}
\label{subsec:tpe}
TPE is the protected text that can silence the LLMs, i.e., prevents them from generating any response and simply terminates the current conversation. To construct TPE, We can formalize the objective of constructing TPE as finding the minimum value of a loss function. Since the characteristic of TPE, an intuitive loss function would be the expected length of the model's response. 

For the input $\mathcal{P}$, let the answer that selected end token in the first round of sampling be $r_e=[\text{end token}]$, $R_{\text{remain}} = R_P - r_e$. Then, we can define this loss function as:
\begin{equation}\label{eq:loss-tpe}
\begin{split}
\mathcal{L}_{TPE}(\mathcal{P}) &= \sum_{r \in R_\mathcal{P}} p(r|\mathcal{P}) \cdot \emph{len}(r) \\
&= p(r_e|\mathcal{P}) \cdot 1 + \sum_{r \in R_{\text{remain}} } p(r|\mathcal{P}) \cdot \emph{len}(r),
\end{split}
\end{equation}
where $\emph{len}(r)$ denotes the number of tokens in $r$.

It is a challenging problem to find a $\mathcal{P}$ that minimizes $\mathcal{L}_{TPE}$ in Equation~\ref{eq:loss-tpe}. However, we notice that by maximizing $p(r_e|\mathcal{P})$, $\mathcal{L}_{TPE}$ can reach its minimum value of $1$.
Therefore, we can transform the problem into optimizing $p(r_e|\mathcal{P})$ to achieve the maximum value, and the final loss function can be represented as:
\begin{equation}\label{eq:Ep-loss}
\mathcal{L}_{TPE}(\mathcal{P}) = -log(p(r_{e}|\mathcal{P})),
\end{equation}
and then we can convert the goal of constructing TPE into an optimization problem:
\begin{equation}\label{eq:opt}
\begin{split}
\underset{\mathcal{P}' \in constraint(\mathcal{V})^{\emph{len}(\mathcal{P})}}{\text{arg min}}  \mathcal{L}_{TPE}(\mathcal{P}),
\end{split}
\end{equation}
where ``$constraint(\mathcal{V})$" refers to the constraint imposed on the available tokens for selection.


\begin{figure*}[t]
    \centering
    \includegraphics[width=17.5cm]{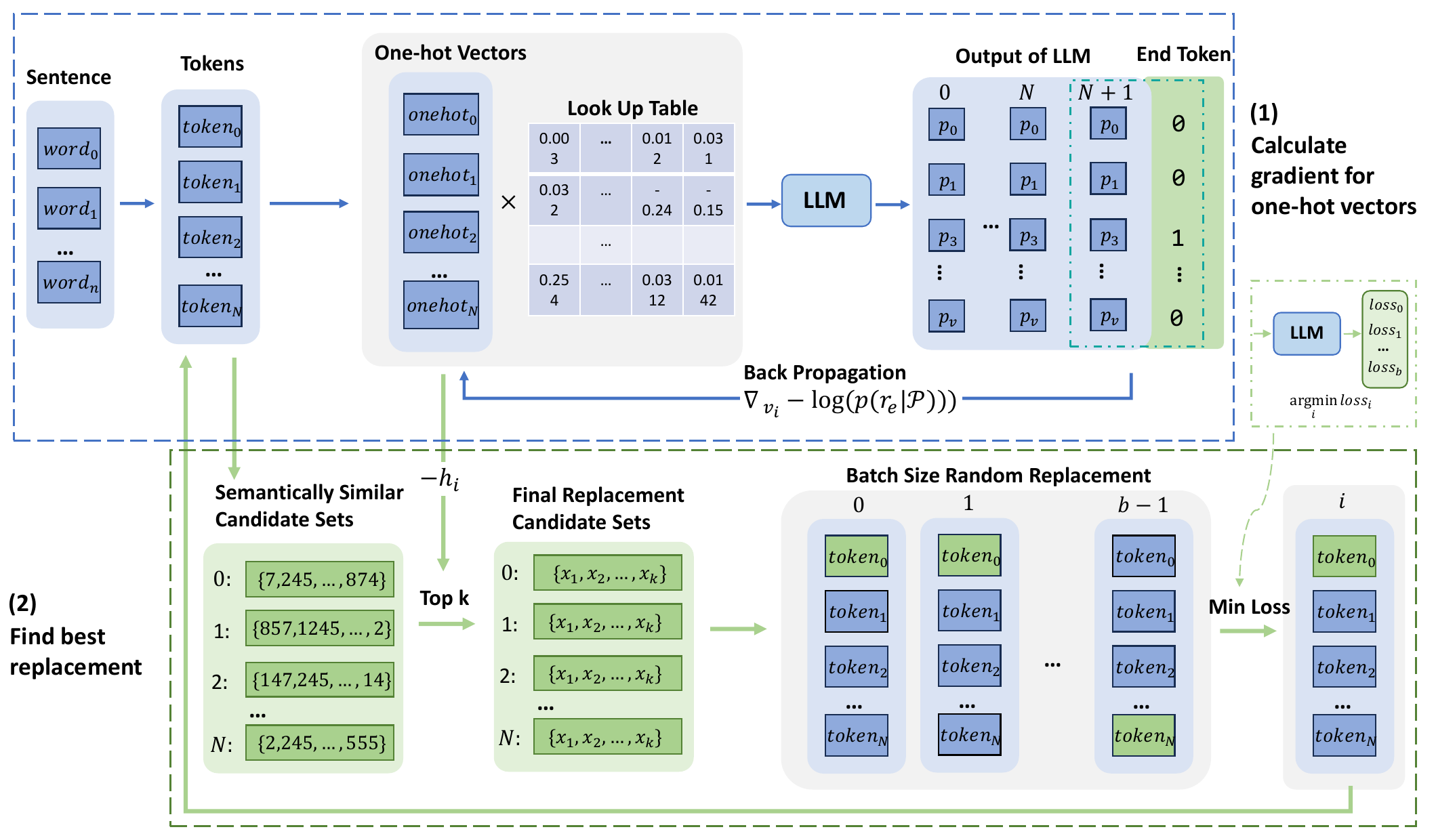}
    \caption{\textbf{The overview of Super Tailored Protection. }(1) Calculate gradient for one-hot vectors: Convert the text to be protected into a one-hot vector representation. Input this into the LLM and utilize the probability distribution of the predicted N+1th token and the end token's probability distribution to compute the loss function. Calculate the gradient and propagate it backward. (2) Find the best replacement: Initially, generate a semantically similar candidate set for each token in the text to be protected using neighboring tokens from the embedding layer. Then, take the results from step 1 to construct the final replacement candidate set from the semantically similar candidate set. Lastly, randomly select and identify the best replacement as the starting text for the next iteration. }
    \label{fig:alg}
\end{figure*}

\subsection{Super Tailored Protection}
\label{subsec:method}
 With the formalized objective of constructing TPE, in this section, we will introduce an effective and stealthy method called Super Tailored Protection (STP) to achieve this.

Figure~\ref{fig:alg} illustrates the overview of STP. The STP method comprises two modules. In the first module, we represent the text to be protected using one-hot vectors, define the loss function $\mathcal{L}_{TPE}$, and compute gradients of one-hot vectors. In the second module, we construct suitable replacement candidate sets, referred to as $constraint(\mathcal{V})$, using gradients that we have computed in the first module. Then, we utilize greedy search to identify the optimal replacements that minimize the loss function. The detailed process is shown below.

\subsubsection{Representation of the prompt using one-hot vectors}
\label{sec:onehot}
Given the prompt to be optimized, denoted as \( \mathcal{P} \), let each token composing \( \mathcal{P} \) be denoted as \( \mathcal{P}_i \). We represent \( \mathcal{P}_i \) as a one-hot vector \( v_i \in \mathbb{R}^V \)
\begin{equation}
\begin{split}
\mathcal{P}=[v_1M_e,v_2M_e,...,v_lM_e].
\end{split}
\end{equation}

\subsubsection{Defining loss function}
\label{sec:loss}
We define the loss function as the cross-entropy between the probability distribution of the first token predicted by the LLM and the probability distribution where the end token has a probability of 1. Specifically, we utilize the output of the LLM logits layer $g$ and the one-hot vector of the end token $v_{\text{end}}$ for computation, i.e., 
\begin{equation}
\text{loss}=H(g,v_{\text{end}}),
\end{equation}

It is worth noting that selecting different loss functions enables the STP method to achieve diverse objectives, showcasing the algorithm's versatility.

\subsubsection{Gradient backpropagation}
\label{sec:back}
Computing the gradient of the one-hot vector corresponding to \(p_i\),
\begin{equation}
\begin{split}
h_i = -\nabla_{v_i} \text{loss}\in \mathbb{R}^V.
\end{split}
\end{equation}
Each dimension of \(h_i\) corresponds to a token in \(\mathcal{V}\), denoted as \(h_i[j]\), where \(j \in \{1, 2, \ldots, V\}\). A smaller \(h_i[j]\) indicates that replacing \(\mathcal{P}_i\) with \(token_j\) would have a larger impact on the loss function, making it converge faster.

\begin{algorithm}[h!]
    \caption{Super Tailored Protection}
    \label{alg:Truncation}
    \DontPrintSemicolon
    \KwIn{Original Prompt $\mathcal{P}$, Iterations $T$, Loss Function $\mathcal{L}$, Batch Size $B$}
    \KwOut{Optimized prompt $\mathcal{P}$}
    \SetKwProg{Fn}{repeat}{}{}
    \Fn{T times}{
        $\text{loss} = \mathcal{L}(\mathcal{P})$\;
        \For{$i=1,...,len(\mathcal{P})$}{
            $h_i = -\nabla_{v_i} \text{loss}$\;
            $N_i=N(\mathcal{P}_i)$\;
            $S_i=Top-k(N_i)$\;
        }
        $\text{len}_{\text{part}}=\frac{B}{len(\mathcal{P})}$\;
        \For{$b=1,...,B$}{
        \uIf{$B > \text{len}(\mathcal{P})$}{$ i=\lceil\frac{b}{\text{len}_{\text{part}}}\rceil$}
        \uElse{
            \Repeat{$i \neq \text{previous } i$}{
                    $i = \text{random}(1, len(\mathcal{P}))$\;
                }
        }
            $\tilde{\mathcal{P}}^{(b)}=\mathcal{P}$\;
            $\tilde{\mathcal{P}}_i^{(b)}=Uniform(S_i)$\;
        }
        $\mathcal{P}=\tilde{\mathcal{P}}^{(b^*)}, \text{ where } b^* = \arg\min_b \mathcal{L}(\tilde{\mathcal{P}}^{(b)})$\;
    }
    \Return $\mathcal{P}$\;
\end{algorithm}

\subsubsection{Construction of semantically similar candidate sets}
\label{sec:similar}
To find semantically similar tokens, we will utilize embeddings~\cite{alzantot2018generating} to find tokens close in the embedding layer.  For each \( \mathcal{P}_i \), select \( n \) closest tokens from $\mathcal{V}$ to construct a semantically similar candidate set. Specifically, We first represent all tokens in the dictionary $\mathcal{V}$ as embedding vectors and normalize them using the $\ell_2$ norm to obtain a new set $\mathcal{V}^{'}$. For token $\mathcal{P}_i$, we perform the same operation, then perform dot products with all vectors in $\mathcal{V}^{'}$, and select the n tokens with the largest results as the set of semantically similar tokens $N_i$.

\subsubsection{Construction of final replacement candidate sets}
\label{sec:final}
To ensure that the replacement maintains similarity with the protected text while causing the loss function to decrease, our final replacement set is selected from within $N_i$ by $h_i$. Specifically, For each token \(token_j \in N_i\), sort them by the \(h_i[j]\) values in descending order. Choose the top \(k\) tokens as the final replacement set, denoted as \(S_i = \text{Top-k}(N_i)\).

\subsubsection{Random replacement and greedy search}
\label{sec:greedy}
In order to accommodate longer lengths of protected text, we employ a combination of random replacement and greedy search to find an optimized prompt. The specific approach is outlined as follows.
In each iteration, repeat \( \mathcal{P} \) \emph{batch size} times and we can obtain an initial set $I=\{\tilde{\mathcal{P}}^{1},\tilde{\mathcal{P}}^{2},...,\tilde{\mathcal{P}}^{\emph{batch size}}\}$, $|I|=\emph{batch size}$. Next, we need to construct a new optimized prompt set, $I^{'}$. Each $\tilde{\mathcal{P}}^{i} \in I$ needs to change the token in one position compared with the original \( \mathcal{P} \) to construct it. The specific method is as follows: 

First, if $batch\ size > \text{len}(\mathcal{P})$, divide $I$ into \( \text{len}(\mathcal{P}) \) parts, $I_1, I_2, \ldots, I_{\text{len}(\mathcal{P})}$, each part corresponding to one changed position $i$. If $batch\ size \leq \text{len}(\mathcal{P})$, randomly select a unique position $i$ for each $\tilde{\mathcal{P}}^{j} \in I$.

Second, Randomly select tokens from \( S_i \) for these positions to perform random replacements, which reduces time consumption for long protected text. Specifically, for $\tilde{\mathcal{P}} \in I_i$, let
\begin{equation}
\tilde{\mathcal{P}}_i=Uniform(S_i).
\end{equation}

Third, Compute the minimum loss replacement among these prompts in each iteration to obtain the new prompt \( \mathcal{P} \). Repeat this process for a specified number of iterations.

The steps described above are presented in Algorithm~\ref{alg:Truncation}.


\section{Experiments and Evaluation}
\label{sec:exp}

\begin{table*}[tb]
\caption{ Result of Truncation Protection Example}
\label{tab:all}
\centering
\begin{tabular}{@{}ccc|ccccccccc@{}}
\toprule
\multirow{2}{*}{Model} & \multirow{2}{*}{Metrics} & \multirow{2}{*}{Method} & \multicolumn{9}{c}{Vicuna Dataset} \\
 & & & Writing & Roleplay & Common-sense & Fermi & Counterfactual & Coding & Math & Generic & Knowledge \\ \midrule
\multirow{9}{*}{Vicuna}& \multirow{3}{*}{\(\gamma\)} & STP & \textbf{0.27} & \textbf{0.26} & \textbf{0.21} & 0.21 & \textbf{0.36} & 0.37 & 0.46 & \textbf{0.37} & \textbf{0.24}\\
& & PEZ & 0.33 & 0.31 & 0.30 & \textbf{0.20} & \textbf{0.36} & \textbf{0.36} & \textbf{0.44} & 0.49 & 0.32\\
& & GBDA & 0.84 & 1.00 & 0.92 & 0.90 & 1.00 & 0.97 & 2.04 & 0.95 & 0.86\\  
\cmidrule{2-12}

&\multirow{3}{*}{\(\eta\)} & STP & 0.73 & \textbf{0.73} & \textbf{0.78} & 0.76 & 0.66 & 0.74 & 0.76 & \textbf{0.66} & \textbf{0.76}\\
& & PEZ & \textbf{0.75} & \textbf{0.73} & 0.72 & \textbf{0.78} & \textbf{0.71} & \textbf{0.76} & \textbf{0.79} & 0.62 & 0.69\\
& & GBDA & 0.50 & 0.50 & 0.50 & 0.51 & 0.51 & 0.52 & 0.49 & 0.48 & 0.50\\
\cmidrule{2-12}

&\multirow{3}{*}{PSR} & STP & \textbf{0.97} & \textbf{0.95} & \textbf{0.88} & \textbf{1.00} & \textbf{0.63} & \textbf{0.79} & \textbf{0.99} & \textbf{0.79} & \textbf{0.88}\\
& & PEZ & 0.05 & 0.05 & 0.07 & 0.05 & 0.08 & 0.03 & 0.06 & 0.07 & 0.10\\
& & GBDA & 0.00 & 0.00 & 0.00 & 0.00 & 0.00 & 0.00 & 0.00 & 0.00 & 0.00\\ \midrule

\multirow{9}{*}{LLaMA}& \multirow{3}{*}{\(\gamma\)} & STP & \textbf{0.26} & \textbf{0.31} & \textbf{0.24} & \textbf{0.21} & 0.44 & 0.40 & \textbf{0.28} & \textbf{0.37} & \textbf{0.27}\\
& & PEZ & 0.35 & 0.38 & 0.41 & 0.36 & \textbf{0.34} & \textbf{0.37} & 0.74 & 0.50 & 0.46\\
& & GBDA & 0.86 & 1.00 & 0.91 & 0.92 & 0.98 & 0.92 & 1.86 & 0.92 & 0.88\\ 
\cmidrule{2-12}
&\multirow{3}{*}{\(\eta\)} & STP & \textbf0.73 & \textbf0.71 & \textbf{0.74} & \textbf{0.75} & 0.63 & 0.69 & \textbf{0.69} & \textbf{0.69} & \textbf{0.68}\\
& & PEZ & \textbf{0.73} & 0.69 & 0.66 & 0.69 & \textbf{0.73} & \textbf{0.72} & 0.65 & 0.61 & 0.64\\
& & GBDA & 0.50 & 0.49 & 0.49 & 0.50 & 0.50 & 0.52 & 0.49 & 0.50 & 0.50\\
\cmidrule{2-12}
&\multirow{3}{*}{PSR} & STP & \textbf{0.53} & \textbf{0.46} & \textbf{0.41} & \textbf{0.65} & \textbf{0.22} & \textbf{0.39} & \textbf{0.44} & \textbf{0.24} & \textbf{0.47}\\
& & PEZ & 0.05 & 0.04 & 0.03 & 0.02 & 0.03 & 0.01 & 0.03 & 0.04 & 0.02\\
& & GBDA & 0.00 & 0.00 & 0.00 & 0.00 & 0.00 & 0.00 & 0.00 & 0.00 & 0.00\\ \midrule

\multirow{9}{*}{Guanaco}& \multirow{3}{*}{\(\gamma\)} & STP & \textbf{0.29} & \textbf{0.33} & \textbf{0.26} & \textbf{0.25} & 0.43 & 0.39 & \textbf{0.44} & \textbf{0.41} & \textbf{0.33}\\
& & PEZ & 0.36 & 0.41 & 0.35 & 0.33 & \textbf{0.31} & \textbf{0.26} & 0.68 & 0.46 & 0.41\\
& & GBDA & 0.84 & 0.98 & 0.89 & 0.88 & 0.98 & 0.91 & 1.79 & 0.91 & 0.87\\ 
\cmidrule{2-12}
&\multirow{3}{*}{\(\eta\)} & STP & 0.71 & 0.68 & \textbf{0.72} & \textbf{0.74} & 0.68 & 0.68 & \textbf{0.70} & 0.64 & \textbf{0.68}\\
& & PEZ & \textbf{0.72} & \textbf{0.70} & 0.69 & 0.71 & \textbf{0.70} &\textbf{0.76} & 0.67 & \textbf{0.65} & 0.65\\
& & GBDA & 0.50 & 0.50 & 0.50 & 0.50 & 0.50 & 0.52 & 0.50 & 0.49 & 0.49\\
\cmidrule{2-12}
&\multirow{3}{*}{PSR} & STP & \textbf{0.56} & \textbf{0.53} & \textbf{0.51} & \textbf{0.67} & \textbf{0.27} & \textbf{0.22} & \textbf{0.70} & \textbf{0.45} & \textbf{0.60}\\
& & PEZ & 0.04 & 0.03 & 0.04 & 0.03 & 0.03 & 0.01 & 0.05 & 0.03 & 0.04\\
& & GBDA & 0.00 & 0.01 & 0.00 & 0.00 & 0.02 & 0.01 & 0.01 & 0.01 & 0.00\\ \bottomrule

\end{tabular}
\end{table*}

\begin{figure*}[h!]
    \centering
    \includegraphics[width=17.5cm]{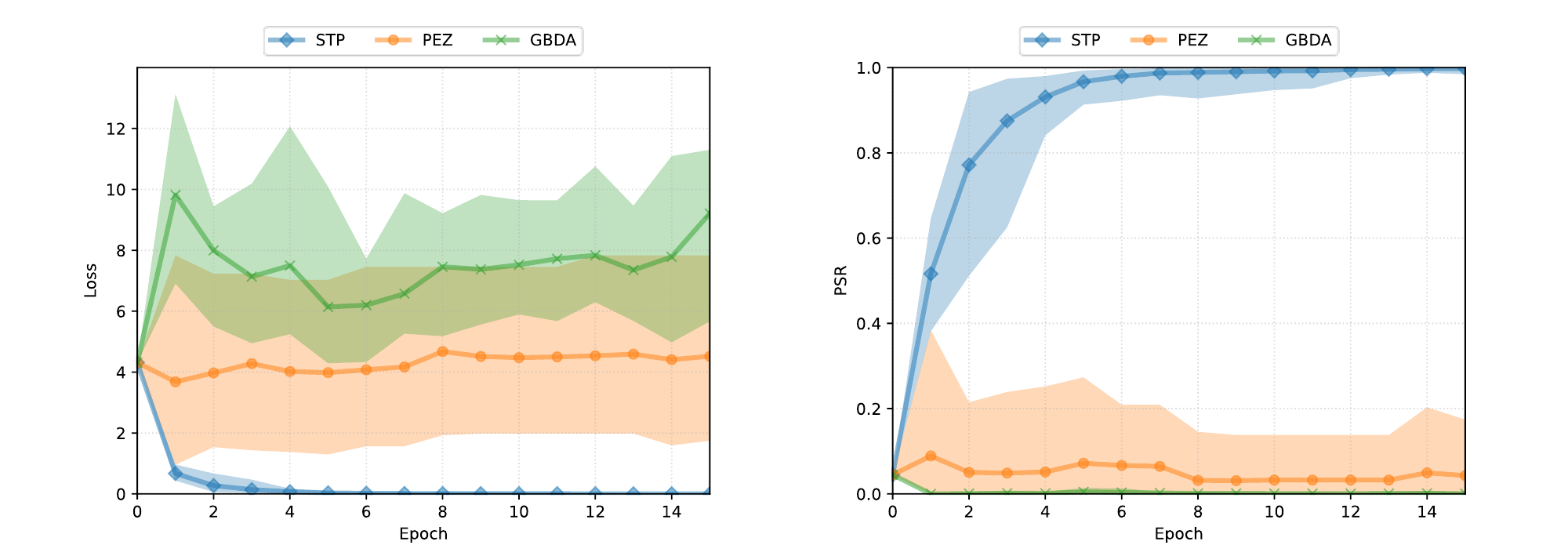}
    \caption{\textbf{The convergence results of Loss and PSR for PEZ, GBDA, and STP methods.} It can be observed that our proposed approach shows faster convergence of loss and higher efficiency when it comes to constructing TPE. It is worth noting that the initial value for optimization in all three methods was set as the original prompt, and the initial steep increase in loss for GBDA is attributed to its deviation from the initial prompt in the first round, mainly due to the introduction of Gumbel-Softmax.
    }
    \label{fig:loss-PSR}
\end{figure*}

\subsection{Setup}
\label{sec:setup}
\subsubsection{Dataset}
In theory, STP does not impose specific requirements on the theme or content of the prompt itself. To comprehensively validate the effectiveness of the defense, we selected 80 prompts across 9 different categories from Vicuna official website~\cite{chiang2023vicuna}, which include writing, roleplay, common-sense, fermi, counterfactual, coding, math, generic, and knowledge. We denote this data set as the Vicuna dataset.

In addition, we selected a set of texts from the novel \emph{Warden} with varying lengths to verify the effectiveness of the text protection method on different text lengths. Specifically, we constructed eight sets of texts of different lengths, each of which is about 40, 80, 120, 160, 200, 240, 280, and 320 tokens long. Each group has 10 texts, totaling 80 texts. We denote this data set as the Novel dataset.

\subsubsection{Model}
To demonstrate the effectiveness of the STP, we conducted experiments on three transformer architecture models. These models are LLaMA~\cite{touvron2023LLaMA}, Vicuna v1.3~\cite{chiang2023vicuna}, and Guanaco~\cite{dettmers2023qlora} in the 7B version. The training of the last two models was built upon the LLaMA model. Specifically, Vicuna was fine-tuned on LLaMA by SFT, while Guanaco was fine-tuned on LLaMA by QLoRA.

Because constructing TPE using the STP method requires model parameters and the transferability of TPE relies on similar model architectures, we did not conduct experiments on non-open-source GPT series models in our study.
\subsubsection{Perparameters}
In this paper, the main parameters are the batch size and the number of elements in the sets $N_i$ and $S_i$, which are all related to the size of the search space, and the latter two are also related to the concealment of the TPE. After finding a trade-off, we selected \(\emph{batch size}=1024\), \(|N(i)|=10\), and \(|S(i)|=5\) to achieve better results. Additionally, we set the number of epochs to \(T=15\). More epochs imply a higher probability of selecting the end token but also result in increased computational overhead.
\subsubsection{Metrics}
We propose three metrics to measure the effectiveness of text protection against LLMs. These metrics are the Character Replacement Ratio \(\gamma\), Semantic Preservation \(\eta\), and the Success Rate of Truncation Protection (PSR). 
\begin{enumerate}[label=\arabic*.]
    \item \(\gamma\), the Character Replacement Ratio, measures the minimum number of characters changed to achieve a certain level of truncation protection. A smaller \(\gamma\) value indicates better concealment because fewer characters are altered. We calculate \(\gamma\) using the Vladimir Levenshtein edit distance~\cite{levenshtein1966binary} divided by the original sentence's character length.
    \item \(\eta\), Semantic Preservation, quantifies the semantic similarity between two sentences before and after token replacement. A higher \(\eta\) value suggests that truncation protection has a smaller impact on the sentence's meaning. We utilized the cosine similarity of sentences~\cite{cer2018universal} to this metric.
    \item PSR, Success Rate of Truncation Protection, represents the effectiveness of truncation protection. We define PSR as the softmax score corresponding to the end token, i.e.,
    \begin{equation}
    \text{PSR}= \frac{e^{z_{end}}} {\sum_{i=1}^V e^{z_i}},
    \end{equation}
    where $z_i$ represents the value in the model's logits layer corresponding to $token_i$.
    A higher PSR indicates a more successful protection effectiveness.
\end{enumerate}

These metrics help evaluate the quality of text protection and its impact on both the text's semantics and the extent to which it effectively truncates the model's output.

\subsubsection{Baseline}
We used GBDA~\cite{guo2021gradient} and PEZ~\cite{wen2023hard} methods as baselines. Equation~\ref{eq:Ep-loss} presents the optimization objective of STP. For baseline, we employed both PEZ and GBDA to achieve this optimization objective. The optimized initial value is set to the text to be protected. Then, we computed the loss function in Equation~\ref{eq:Ep-loss} and utilized PEZ and GBDA algorithms individually to optimize the prompt. It's important to note that while the GBDA method offers constraints on the concealment of adversarial examples, it requires a specific inference model. Hence, in this paper, we didn't impose concealment constraints on GBDA. Additionally, to prevent gradient explosions when using these two methods, we used the Adam optimizer with values of epsilon (eps) $1e-5$.
\subsection{Evaluation}
\subsubsection{White Box}
\label{sec:white-box}
\textbf{The Comprehensiveness of the Truncation Protection Example. }Table~\ref{tab:all} shows the results of TPE constructed by STP, PEZ, and GBDA on nine categories of prompts from the Vicuna dataset on LLaMA, Vicuna, and Guanaco. We applied 15 rounds of replacements to each of the 80 prompts, with each round signifying one token replacement of the target prompt.

The experimental results indicate that STP exhibits favorable protective effects for prompts of varying categories and lengths. However, among the nine different categories of prompts tested, prompts related to Counterfactual and Math, due to their shorter lengths, experienced greater disruption from replacement, leading to relatively poorer concealment. Compared to STP, PEZ and GBDA show little to no effectiveness in constructing TPE.

To illustrate the superiority of the STP method in constructing TPE, we selected ten prompts from the ``Writing" category in the Vicuna dataset. We conducted TPE construction using the STP method, PEZ, and GBDA methods on Vicuna, and the final results are depicted in Figure~\ref{fig:loss-PSR}.

Furthermore, in our experiments, we observed that a significant portion of the replacement operations involved changes in capitalization, such as transforming ``you" to ``You" or substitutions with visually similar or identical characters but different encodings, like changing ``a" to ``а". This ensured the concealment of TPE and highlighted the advantage of using tokens rather than words as the smallest replacement units.Table~\ref{tab:TPE-exampless} displays some examples of TPE.

Moreover, due to the greedy search principle followed by the STP method, replacements that had a considerable impact on the outcome were often in fixed positions within sentences. Consequently, many rounds of replacement were focused on the same location, reinforcing the concealment of TPE.

It is important to note that conducting 15 rounds of construction did not achieve convergence in the loss function for most prompts. Therefore, conducting more rounds of construction for a single prompt should yield better protective results, but this might also lead to greater alterations in the prompt itself.
\begin{table}[h!]
\caption{ Examples of TPE on Vicuna while PSR $\geq$ 0.95}
\label{tab:TPE-exampless}
\centering
\begin{tabular}{@{}p{4cm}p{4cm}@{}}
\toprule
  \makecell[c]{Original Text} & \makecell[c]{Protected Text}\\ \midrule
 \textcolor{blue}{Can y}ou help me write a resignation letter to my current employer, while leaving on good terms \textcolor{blue}{and} expressing grat\textcolor{blue}{itude} for the opportunities provided\textcolor{blue}{?} &
 \textcolor{red}{ShouldY}ou help me write a resignation letter to my current employer, while leaving on good terms \textcolor{red}{but} expressing grat \textcolor{red}{attitude} for the opportunities provided\textcolor{red}{？} \\ \midrule
 \textcolor{blue}{D}raft an apology \textcolor{blue}{email} to \textcolor{blue}{a} customer \textcolor{blue}{w}ho experienced a delay \textcolor{blue}{i}n their order, and provide reassurance that the issue has been \textcolor{blue}{resolved}. &
 \textcolor{red}{Д}raft an apology \textcolor{red}{contact} to \textcolor{red}{а} customer\textcolor{red}{W}ho experienced a delay\textcolor{red}{I}n their order,\textcolor{red}{\textbackslash r} and provide reassurance that the issue has been \textcolor{red}{revol}. \\ \midrule
 Write a comp\textcolor{blue}{elling} product launch announcement email to inform our customers of our new software \textcolor{blue}{solution}.&
 Write a comp \textcolor{red}{sell} product launch announcement email to inform our customers of our new software \textcolor{red}{answer\textdollar?} \\ \midrule
 Write a cap\textcolor{blue}{ti}vating movie \textcolor{blue}{review} for \textcolor{blue}{a} recently released science fiction film, discus\textcolor{blue}{si}ng its plot, characters, and special effects\textcolor{blue}{.} &
 Write a capt\textcolor{red}{v}ivating movie \textcolor{red}{evaluate} for a\textcolor{red}{n} recently released science fiction film, discuss\textcolor{red}{iz}ing its plot, character, and special effects\textcolor{red}{\}\textdollar}. \\ \bottomrule

\end{tabular}
\end{table}

\textbf{The Effectiveness of Truncation Protection for Texts of Different Lengths.} With increasing prompt length, the search space for constructing TPE grows, and the concealment is enhanced under the same number of replacements. To investigate this, we conducted 15 rounds of TPE construction on the prepared Novel dataset. Table~\ref{tab:different-length} shows the effectiveness of STP for texts of different token lengths. As expected, under the same number of rounds, the final effectiveness of TPE remains largely consistent. An improvement was even observed in LLaMA and Guanaco. Simultaneously, the concealment of TPE increases with the length of the text. This result means that our method has a distinctive advantage in protecting long texts and is highly suitable for real-world applications.

\begin{figure}[h!]
    \centering
    \includegraphics[width=8cm]{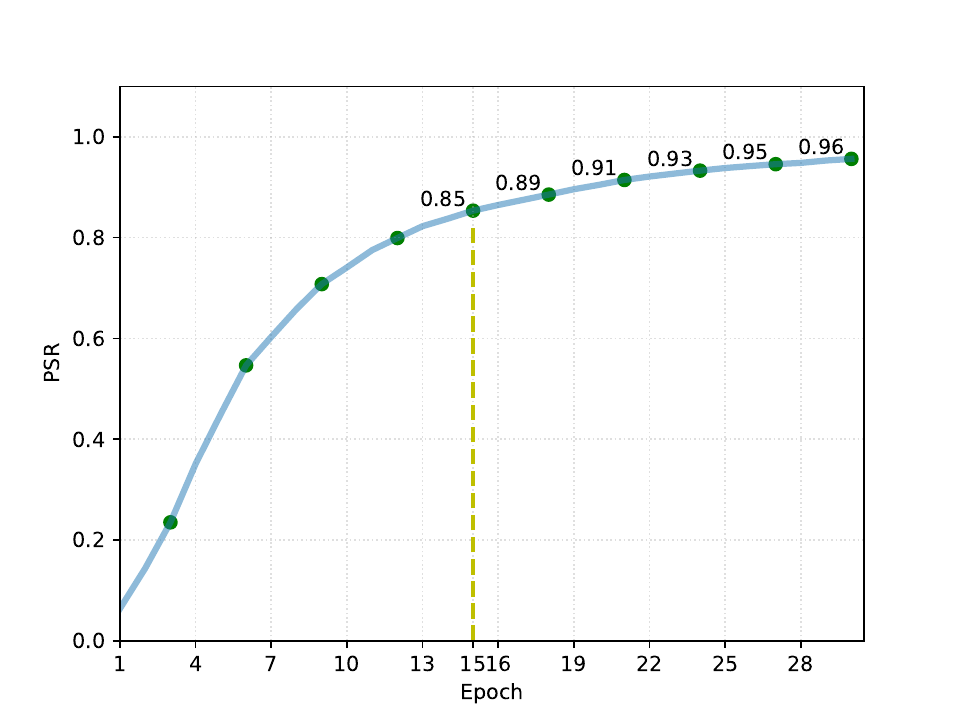}
    \caption{\textbf{Constructing TPE over 30 rounds on text of 120 tokens in the Novel dataset.}}
    \label{fig:example-30epoc}
\end{figure}

As mentioned in section~\ref{sec:white-box}, stronger protection for a single text can be achieved by increasing the number of epochs. In this section, to control variables, we selected texts in 120 tokens from the Novel dataset and conducted 30 epochs of construction on Vicuna. Figure~\ref{fig:example-30epoc} represents the effectiveness of 30 rounds of construction.

\begin{table}[h!]
\caption{ Result of different lengths of text}
\label{tab:different-length}
\centering
\begin{tabular}{@{}cc|ccc@{}}
\toprule
\multirow{2}{*}{Model} & \multirow{2}{*}{Metrics} & \multicolumn{3}{c}{Novel Dataset} \\
 & & 40 tokens & 80 tokens & 120 tokens  \\ \midrule
 \multirow{3}{*}{LLaMA} & \(\gamma\) & 0.18 & 0.09 & 0.06 \\
 & \(\eta\) & 0.75 & 0.85 & 0.88\\
 & PSR & 0.49 & 0.56 & 0.69 \\ \midrule
 \multirow{3}{*}{Vicuna}& \(\gamma\) & 0.19 & 0.09 & 0.06 \\
 & \(\eta\) & 0.76 & 0.84 & 0.89 \\
 & PSR & 0.83 & 0.81 & 0.81 \\ \midrule
 \multirow{3}{*}{Guonaco} & \(\gamma\) & 0.18 & 0.09 & 0.07 \\
 & \(\eta\) & 0.76 & 0.84 & 0.86\\
 & PSR & 0.57 & 0.66 & 0.67 \\ \bottomrule

\end{tabular}
\end{table}

\begin{table*}[t]
\caption{ Truncation Protection Examples with added prefixes}
\label{tab:pre}
\centering
\begin{tabular}{@{}cc|ccc|ccc|ccc@{}}
\toprule
\multirow{2}{*}{Model} & \multirow{2}{*}{Metrics} & \multicolumn{3}{c|}{$\rm Prefix_1$}  & \multicolumn{3}{c|}{$\rm Prefix_2$} & \multicolumn{3}{c}{$\rm Prefix_3$}\\
 & & 40 tokens & 80 tokens & 120 tokens & 40 tokens & 80 tokens & 120 tokens & 40 tokens & 80 tokens & 120 tokens \\ \midrule
\multirow{3}{*}[0.6ex]{Vicuna}& PSR & 0.84 & 0.82 & 0.82 & 0.84 & 0.82 & 0.82 & 0.84 & 0.82 & 0.82\\
& PSR* & 0.48 & 0.39 & 0.46 & 0.46 & 0.39 & 0.46 & 0.43 & 0.38 & 0.43\\ \midrule

\multirow{3}{*}[0.6ex]{LLaMA} & PSR & 0.49 & 0.56 & 0.69 & 0.49 & 0.56 & 0.69 & 0.49 & 0.56 & 0.69\\
& PSR* & 0.11 & 0.21 & 0.31 & 0.12 & 0.19 & 0.31 & 0.12 & 0.21 & 0.37\\ \midrule

  \multirow{3}{*}[0.6ex]{Guanaco} & PSR & 0.57 & 0.66 & 0.68 & 0.57 & 0.66 & 0.68 & 0.57 & 0.66 & 0.68\\
& PSR* & 0.14 & 0.23 & 0.27 & 0.12 & 0.27 & 0.29 & 0.19 & 0.28 & 0.36\\ \bottomrule

\end{tabular}
\end{table*}

\begin{table*}[t]
\caption{ Truncation Protection Examples with added suffixes}
\label{tab:suf}
\centering
\begin{tabular}{@{}cc|ccc|ccc|ccc@{}}
\toprule
\multirow{2}{*}{Model} & \multirow{2}{*}{Metrics} & \multicolumn{3}{c|}{$\rm Suffix_1$}  & \multicolumn{3}{c|}{$\rm Suffix_2$} & \multicolumn{3}{c}{$\rm Suffix_3$}\\
 & & 40 tokens & 80 tokens & 120 tokens & 40 tokens & 80 tokens & 120 tokens & 40 tokens & 80 tokens & 120 tokens  \\ \midrule
\multirow{3}{*}[0.6ex]{Vicuna}& PSR & 0.84 & 0.82 & 0.82 & 0.84 & 0.82 & 0.82 & 0.84 & 0.82 & 0.82\\
& PSR* & 0.29 & 0.39 & 0.37 & 0.25 & 0.33 & 0.39 & 0.53 & 0.51 & 0.56\\ \midrule

\multirow{3}{*}[0.6ex]{LLaMA} & PSR & 0.49 & 0.56 & 0.69 & 0.49 & 0.56 & 0.69 & 0.49 & 0.56 & 0.69\\
& PSR* & 0.11 & 0.15 & 0.12 & 0.15 & 0.20 & 0.19 & 0.21 & 0.28 & 0.25\\ \midrule

 \multirow{3}{*}[0.6ex]{Guanaco} & PSR & 0.57 & 0.66 & 0.68 & 0.57 & 0.66 & 0.68 & 0.57 & 0.66 & 0.68\\
& PSR* & 0.21 & 0.15 & 0.22 & 0.21 & 0.18 & 0.27 & 0.38 & 0.40 & 0.52\\ \bottomrule

\end{tabular}
\end{table*}

\textbf{Real-World Scenarios. }In real-world scenarios, adversaries often add prefixes and suffixes to texts, such as ``Please summarize the following text: [exploited text]" or ``[exploited text]. Please summarize the preceding topic." These are important steps when using LLMs for content generation. Therefore, In this section, we conducted experiments on Vicuna, LLaMA, and Guanaco by adding prefixes ``Summarize following text:" ``Summarize the topic of following text:" ``Please summarize the topic of the following text and rewrite:" and suffixes ``Summarize the preceding text." ``Summarize the topic of the preceding text." ``Please summarize the topic of the preceding text and rewrite." to the TPE on the Novel dataset. These are denoted as $\rm prefix_1$, $\rm prefix_2$, $\rm prefix_3$, $\rm suffix_1$, $\rm suffix_2$, and $\rm suffix_3$, respectively. 



The experimental results are in Table~\ref{tab:pre} and Table~\ref{tab:suf}. The PSR and PSR* in the table, respectively, represent the Protection Success Rate of TPE before and after adding a prefix or suffix. Even though we did not specifically optimize the protection for these prefixes and suffixes, the PSR remains at a relatively high level. Furthermore, as the length of the text to be protected increases, the impact of adding prefixes and suffixes diminishes on the protective effect. Additionally, longer prefixes and suffixes have a relatively minor effect on the protective outcome. This suggests that STP exhibit superior performance in tasks involving long texts and complex prefixes/suffixes, laying the groundwork for their scalability.

\subsubsection{Transferability}
In this section, we conducted transferability experiments of TPE. Specifically, we utilized the results optimized on the Vicuna, Llama, and Guanaco models interchangeably across all three models. The experimental results are presented in Table~\ref{tab:transfer-novel} and Table~\ref{tab:transfer}. Here, A$\rightarrow$B denotes the transfer of optimized results from model A to model B. PSR represents the Protection Success Rate of the TPE constructed on the original model, while PSR* signifies the transferred Protection Success Rate.

\begin{table}[h!]
\caption{Transferability of Truncation Protection Examples on Novel dataset}
\label{tab:transfer-novel}
\centering
\begin{tabular}{@{}cc|ccc@{}}
\toprule
\multirow{2}{*}{Model} & \multirow{2}{*}{Metrics} & \multicolumn{3}{c}{Novel Dataset} \\
 & & 40tokens & 80tokens & 120tokens  \\ \midrule
\multirow{3}{*}[0.5ex]{\makecell[c]{Vicuna$\rightarrow$LLaMA}}& PSR & 0.49 & 0.56 & 0.69 \\
 & PSR* & 0.13 & 0.18 & 0.16 \\ \midrule
\multirow{3}{*}[0.5ex]{\makecell[c]{LLaMA$\rightarrow$Vicuna}} & PSR & 0.83 & 0.81 & 0.81 \\
 & PSR* & 0.10 & 0.19 & 0.41\\ \midrule
 \multirow{3}{*}[0.5ex]{\makecell[c]{LLaMA$\rightarrow$Guanaco}} & PSR & 0.57 & 0.66 & 0.67 \\
 & PSR* & 0.23 & 0.39 & 0.46\\ \midrule
 \multirow{3}{*}[0.5ex]{\makecell[c]{Guanaco$\rightarrow$LLaMA}} & PSR & 0.49 & 0.56 & 0.69 \\
 & PSR* & 0.24 & 0.26 & 0.28\\ \midrule
 \multirow{3}{*}[0.5ex]{\makecell[c]{Vicuna$\rightarrow$Guanaco}} & PSR & 0.57 & 0.66 & 0.67 \\
 & PSR* & 0.13 & 0.25 & 0.27\\ \midrule
 \multirow{3}{*}[0.5ex]{\makecell[c]{Guanaco$\rightarrow$Vicuna}} & PSR & 0.83 & 0.81 & 0.81 \\
 & PSR* & 0.18 & 0.22 & 0.31\\ \bottomrule
\end{tabular}
\end{table}

In the Novel dataset, experimental results show that transferability improves with the increase in text length. However, in the Vicuna dataset, we observed that the TPE optimized for different models exhibited weaker transferability between models. To address this issue, we aggregated the loss functions of different models into a new loss function to construct TPE, hoping to enhance transferability to some extent.

Specifically, for three models $\text{model}_1$, $\text{model}_2$, and $\text{model}_3$, we defined the new loss function as $(\text{loss}_{\text{model}_1}+\text{loss}_{\text{model}_2})/2$, and used this loss function to construct TPE, which is used to attack $\text{model}_3$. The specific experimental results are shown in Table \ref{tab:transfer_up}. Here, AVE represents the average of the results in the first two rows of the table, and AGG represents the aggregating of the other two models.

The experimental results show that the TPE constructed using the aforementioned method exhibits enhanced transferability. Compared to the original method, the average PSR values across various models in the Vicuna dataset have increased by 0.11. Furthermore, from the above experimental results, it can be reasonably inferred that aggregating the loss functions of more models to construct TPE should lead to even better transferability.

\begin{table*}[h]
\caption{Transferability of Truncation Protection Examples on Vicuna dataset}
\label{tab:transfer}
\centering
\begin{tabular}{@{}cc|ccccccccc@{}}
\toprule
\multirow{2}{*}{Model} & \multirow{2}{*}{Metrics} & \multicolumn{9}{c}{Vicuna Dataset} \\
 & & Writing & Roleplay & Common-sense & Fermi & Counterfactual & Coding & Math & Generic & Knowledge \\ \midrule
\multirow{3}{*}[0.5ex]{\makecell[c]{Vicuna$\rightarrow$LLaMA}}& PSR & 0.53 & 0.46 & 0.41 & 0.65 & 0.22 & 0.39 & 0.44 & 0.24 & 0.47\\
& PSR* & 0.07 & 0.10 & 0.05 & 0.10 & 0.05 & 0.07 & 0.09 & 0.05 & 0.08\\ \midrule
\multirow{3}{*}[0.5ex]{\makecell[c]{LLaMA$\rightarrow$Vicuna}} & PSR & 0.97 & 0.95 & 0.88 & 1.00 & 0.63 & 0.79 & 0.99 & 0.79 & 0.88\\
& PSR* & 0.41 & 0.27 & 0.37 & 0.60 & 0.13 & 0.16 & 0.40 & 0.31 & 0.22\\ \midrule
\multirow{3}{*}[0.5ex]{\makecell[c]{LLaMA$\rightarrow$Guanaco}} & PSR & 0.56 & 0.53 & 0.51 & 0.67 & 0.27 & 0.22 & 0.70 & 0.45 & 0.60\\
& PSR* & 0.26 & 0.22 & 0.30 & 0.43 & 0.12 & 0.13 & 0.27 & 0.16 & 0.26\\ \midrule
\multirow{3}{*}[0.5ex]{\makecell[c]{Guanaco$\rightarrow$LLaMA}} & PSR & 0.53 & 0.46 & 0.41 & 0.65 & 0.22 & 0.39 & 0.44 & 0.24 & 0.47\\
& PSR* & 0.18 & 0.20 & 0.20 & 0.19 & 0.09 & 0.12 & 0.17 & 0.11 & 0.26\\ \midrule
\multirow{3}{*}[0.5ex]{\makecell[c]{Vicuna$\rightarrow$Guanaco}} & PSR & 0.56 & 0.53 & 0.51 & 0.67 & 0.27 & 0.22 & 0.70 & 0.45 & 0.60\\
& PSR* & 0.17 & 0.10 & 0.10 & 0.28 & 0.06 & 0.05 & 0.14 & 0.06 & 0.11\\ \midrule
\multirow{3}{*}[0.5ex]{\makecell[c]{Guanaco$\rightarrow$Vicuna}} & PSR & 0.97 & 0.95 & 0.88 & 1.00 & 0.63 & 0.79 & 0.99 & 0.79 & 0.88\\
& PSR* & 0.26 & 0.38 & 0.37 & 0.62 & 0.18 & 0.13 & 0.26 & 0.25 & 0.30\\ \bottomrule
\end{tabular}
\end{table*}

\begin{table*}[h]

\caption{Enhancement of TPE Transferability on the Vicuna Dataset}
\label{tab:transfer_up}
\centering
\begin{tabular}{@{}cc|ccccccccc@{}}
\toprule
\multirow{2}{*}{Model} & \multirow{2}{*}{Metrics} & \multicolumn{9}{c}{Vicuna Dataset} \\
 & & Writing & Roleplay & Common-sense & Fermi & Counterfactual & Coding & Math & Generic & Knowledge \\ \midrule
Guanaco & PSR & 0.56 & 0.53 & 0.51 & 0.67 & 0.27 & 0.22 & 0.70 & 0.45 & 0.60\\ \midrule
LLaMA$\to$Guanaco & PSR* & 0.26 & 0.22 & 0.30 & 0.43 & 0.12 & 0.13 & 0.27 & 0.16 & 0.26\\ \midrule
Vicuna$\to$Guanaco & PSR* & 0.17 & 0.10 & 0.10 & 0.28 & 0.06 & 0.05 & 0.14 & 0.06 & 0.11\\ \midrule
 $\backslash$ & AVE & 0.21 & 0.19 & 0.20 & 0.35 & 0.09 & 0.09 & 0.20 & 0.11 & 0.18\\ \midrule
AGG$\to$Guanaco & PSR* & 0.22$\uparrow$ & 0.28$\uparrow$ & 0.26$\uparrow$ & 0.30 & 0.20$\uparrow$ & 0.16$\uparrow$ & 0.40$\uparrow$ & 0.31$\uparrow$ & 0.40$\uparrow$\\ \midrule\midrule

LLaMA & PSR & 0.53 & 0.46 & 0.41 & 0.65 & 0.22 & 0.39 & 0.44 & 0.24 & 0.47\\ \midrule
Vicuna$\to$LLaMA & PSR* & 0.07 & 0.10 & 0.05 & 0.10 & 0.05 & 0.07 & 0.09 & 0.05 & 0.08\\ \midrule
Guanaco$\to$LLaMA & PSR* & 0.18 & 0.20 & 0.20 & 0.19 & 0.09 & 0.12 & 0.17 & 0.11 & 0.26\\ \midrule
 $\backslash$ & AVE & 0.12 & 0.15 & 0.12 & 0.14 & 0.07 & 0.09 & 0.13 & 0.08 & 0.17\\ \midrule
AGG$\to$LLaMA & PSR* & 0.24$\uparrow$ & 0.25$\uparrow$ & 0.26$\uparrow$ & 0.20$\uparrow$ & 0.15$\uparrow$ & 0.18$\uparrow$ & 0.25$\uparrow$ & 0.21$\uparrow$ & 0.32$\uparrow$\\ \midrule\midrule

Vicuna & PSR & 0.97 & 0.95 & 0.88 & 1.00 & 0.63 & 0.79 & 0.99 & 0.79 & 0.88\\ \midrule
LLaMA$\to$Vicuna & PSR* & 0.41 & 0.27 & 0.37 & 0.60 & 0.13 & 0.16 & 0.40 & 0.31 & 0.22\\ \midrule
Guanaco$\to$Vicuna & PSR* & 0.26 & 0.38 & 0.37 & 0.62 & 0.18 & 0.13 & 0.26 & 0.25 & 0.30\\ \midrule
 $\backslash$ & AVE & 0.33 & 0.32 & 0.37 & 0.61 & 0.15 & 0.14 & 0.33 & 0.28 & 0.26\\ \midrule
AGG$\to$Vicuna & PSR* & 0.41$\uparrow$ & 0.41$\uparrow$ & 0.50$\uparrow$ & 0.48 & 0.28$\uparrow$ & 0.44$\uparrow$ & 0.48$\uparrow$ & 0.44$\uparrow$ & 0.54$\uparrow$\\ \bottomrule

\end{tabular}
\end{table*}

\subsubsection{The Relationship between the Time and Hyperparameters for Constructing a Single TPE}
\label{sec:text_para}
In the preceding sections, we analyzed the practical effectiveness of STP in text protection. In this section, we delve into the relationship between TPE construction time and various parameters. As depicted in Figure~\ref{fig:alg}, the construction time of TPE is primarily associated with the length of the text to be protected, the size of the replacement set, the number of construction rounds, and the batch size. In our experimental setup, STP for a single text on the 7B model using an NVIDIA Quadro RTX8000 GPU takes approximately 12 minutes. We will now investigate the correlation between these parameters and the construction time of TPE.

\textbf{Text length.} The text selected as the protection target is not an adjustable parameter. However, fortunately, due to the parallel computing nature of transformer models, the computational time for texts of different lengths remains relatively consistent. Similar to the experimental settings in Section~\ref{sec:exp}, we conducted 15 epochs of construction for texts in the Novel Dataset on the Vicuna-7B model. The average construction times in 40 tokens, 80 tokens, and 120 tokens text were approximately 12.95 minutes, 13.15 minutes, and 15.18 minutes, respectively.

The experimental results indicate a slight increase in construction time with an increase in the length of the protected text. Nevertheless, this increase remains within an acceptable range.

\textbf{Size of replacement sets.} The time complexity involved in constructing replacement sets is negligible compared to a single forward pass of the model. However, the size of the replacement set significantly influences the convergence rate of the loss function. Qualitatively, a larger semantically similar candidate set \(N_i\) broadens the model's selection scope, providing more efficient options when forming the final replacement candidate set \(S_i\). However, it will also increase the randomness in the algorithm.

\textbf{Number of construction rounds.} Increasing the number of construction rounds enhances the success rate of protection but also extends the time required.

\textbf{Batch size.} Larger batch sizes imply a larger selection space, leading to a greater reduction in the loss function within a single round of construction. However, it also results in increased construction time.

\begin{figure}[t]
    \centering
    \includegraphics[width=8cm]{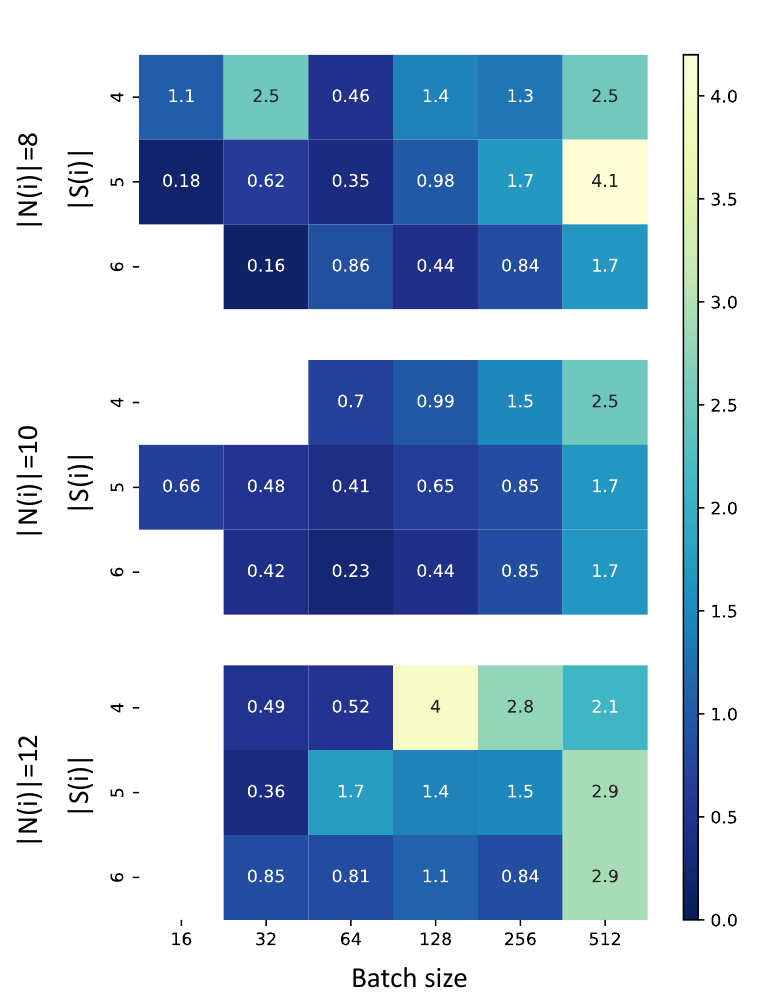}
    \caption{\textbf{The heatmap of the relationship between the replacement set size, batch size, and construction time.} The masked areas indicate scenarios where, even after 100 epochs, the PSR did not reach 0.9, and the unit of the numbers in the figure in minutes.}
    \label{fig:time}
\end{figure}

After discussing the impact of the aforementioned parameters on construction time, in order to maintain semantic similarity, we conducted experiments by selecting a set of data near the original $|N(i)|$ and $|S(i)|$ to find more optimal parameters for reducing the time cost of constructing TPE. Specifically, we set $|N(i)|\in\{8, 10, 12\}$, $|S(i)|\in\{4, 5, 6\}$, and $batch\;size\in\{16, 32, 64, 128, 256, 512\}$. We then selected a text from the Vicuna dataset for experimentation, terminating training when the text's PSR reached or exceeded 90\%, and recorded the time taken for the calculations. The experimental results are depicted in Figure~\ref{fig:time}.

It can be observed that reducing the batch size significantly decreases the running time of STP, with the optimal scenario taking only about 6 seconds. Similarly, adjusting other parameters properly can also significantly reduce the running time of STP when a higher PSR is required.

\subsubsection{Discussion on the Relationship between TPE Construction Time and Text Length}
\begin{figure}[h!]
    \centering
    \includegraphics[width=8cm]{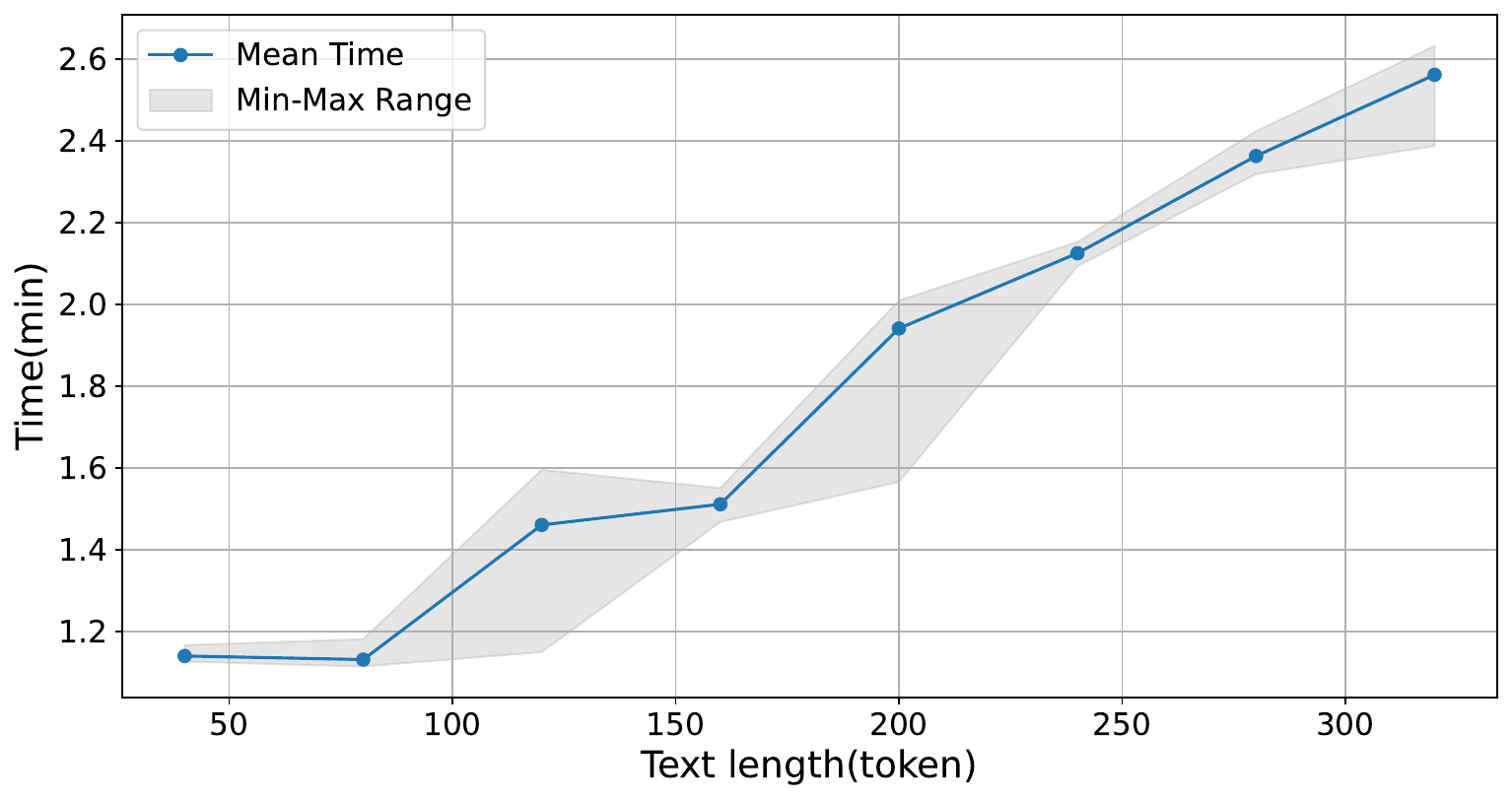}
    \caption{\textbf{The relationship between TPE construction time and text length.}
    }
    \label{fig:time_text_length}
\end{figure}

In this section, we will discuss the relationship between the time required for TPE construction and the length of the text. Specifically, we conducted experiments on texts of lengths 40, 80, 120, 160, 200, 240, 280, and 320 tokens from the Novel dataset. The experimental parameters were set as $|S_i|=5$, $|N_i|=10$, $batch\ size=128$, $T=15$. The experimental results are shown in Figure \ref{fig:time_text_length}. All experiments in this section were completed using an NVIDIA RTX A6000 GPU.

The experimental results indicate that the time cost is approximately linearly related to the length of the text. Furthermore, as discussed in Section \ref{sec:text_para} of the manuscript, adjusting the parameter settings appropriately can further reduce the time cost. Additionally, our text protection method can be applied in offline scenarios without the need for real-time computation, making its application in real-world scenarios feasible.

\subsubsection{Why Truncation Works}
In Section~\ref{subsec:method}, we mentioned that the effectiveness of constructing TPE is attributed to the tendency of dialogues to naturally end. In this section, we emphasize this point and delve deeper into the intrinsic nature of the model.

To ensure clearer contrasts in the model experimental outcomes, we opted for six prompts from the Vicuna dataset as the text examples to be protected, leveraging Vicuna as the target model. Upon inputting these prompts into the model, predictions for the next token were obtained. We selected approximately four to five tokens with probabilities close to the end token as our optimization targets to control variables. Specifically, we focused on the last dimension of the logits layer output, denoted as $output.logits[0,-1]$, and sorted these tokens in descending order. We then chose the three tokens preceding the end token and the three tokens succeeding it as our new optimization targets.

\begin{figure*}[t]
    \centering
    \includegraphics[width=17.5cm]{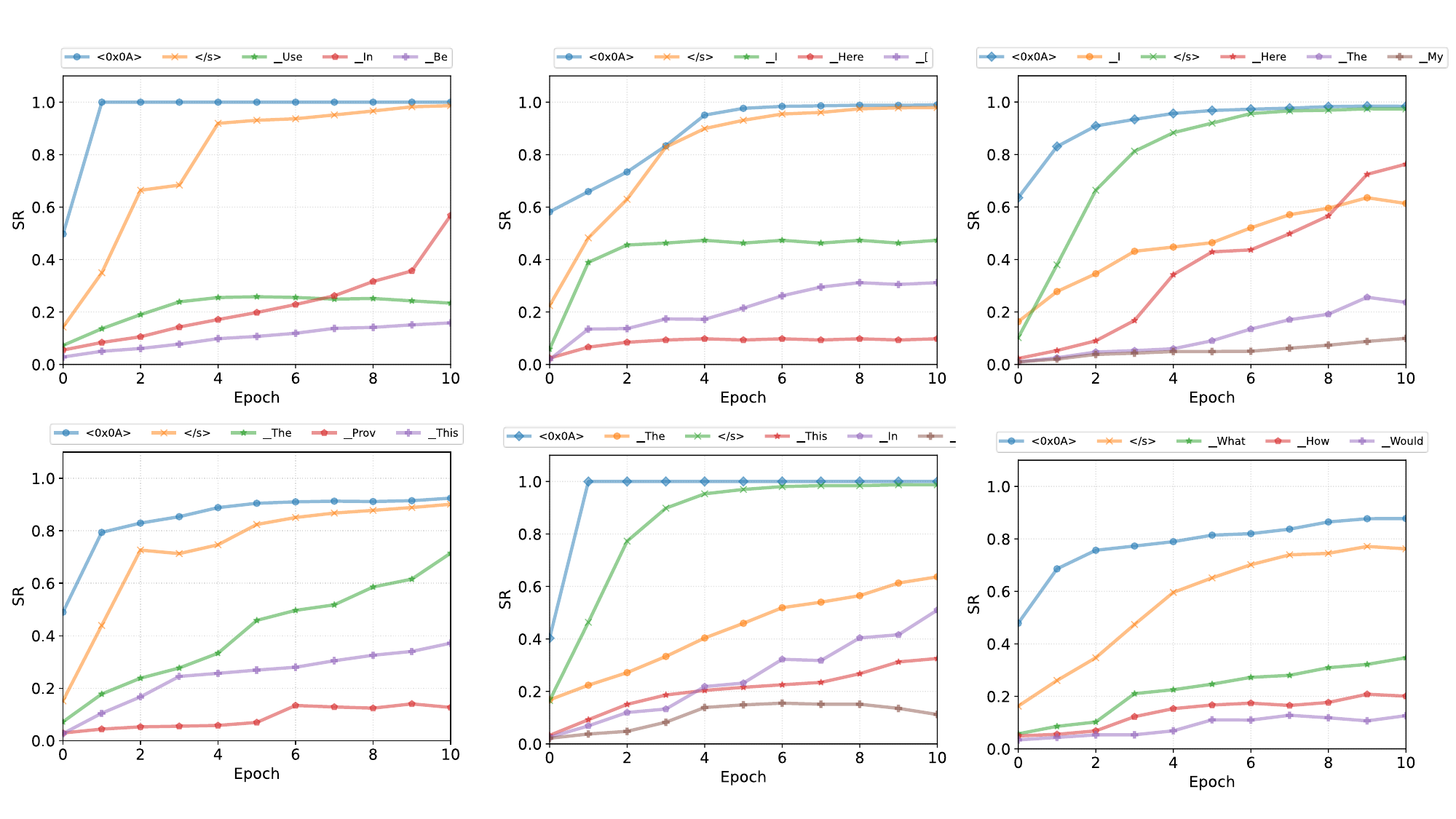}
    \caption{\textbf{The optimization result of six text examples targeting different tokens.} Each line plot corresponds to the optimization results for a specific set of tokens targeted in the text. The horizontal axis Epoch represents the number of optimization rounds, and the vertical axis SR (success rate) represents the probability of the model outputting the token, calculated similarly to PSR. The legend above each plot shows, from left to right, the top five or six tokens with the highest SR rankings before optimization.}
    \label{fig:why-truncation-can}
\end{figure*}

It is noteworthy that in Vicuna, we observed that the end token usually ranks second or third among all tokens. While this high ranking was somewhat unexpected, it did explain the favorable performance of the Vicuna model in our previous experiments. Additionally, in our experiments, the token ranked first consistently corresponded to token ``$\langle 0x0A \rangle $", indexed as token 13 in the dictionary, which is one of the foundational subwords used in the initialization of the vocabulary for the BBPE~\cite{wang2020neural} algorithm.

Subsequently, we optimized the text to increase the probability of the model outputting the first token as the target token, iterating this process for ten rounds while maintaining other experimental settings as outlined in Section~\ref{sec:setup}. The experimental results are depicted in Figure~\ref{fig:why-truncation-can}.

The results indicate that optimizing for the end token and ``$\langle 0x0A \rangle$" as optimization targets were easily achieved, while other tokens were less sensitive to STP's adjustments. Notably, some optimization targets initially exhibited higher values than the end token, gradually being overtaken by the end token across the epochs, highlighting the particularity of the end token. We refer to tokens sensitive to the optimization algorithm as ``sensitive tokens," each corresponding to inherent properties of the model. For instance, the end token indicates a propensity to end dialogues, tokens ranking higher represent the model's compliance with instructions, while tokens like ``$\langle 0x0A \rangle$" correspond to certain biases generated during the model's training. Exploring these biases may unveil deeper security implications.
\section{Discussion}
\label{sec:discussions}
\subsection{Significance of This Work and Future Directions}
The proposed Silent Guardian in this paper represents the first text protection mechanism for LLMs, addressing the security gaps associated with malicious exploitation. As multimodal models and multidimensional models continue to emerge, the protective measures outlined in this paper can be extended to encompass a broader range of generative domains, including audio, images, videos, and beyond. It is anticipated that this extension will have profound implications for the field of generative large models.

\subsection{STP Method}
As shown in Figure~\ref{fig:example}, the STP method ensures good concealment while allowing for rapid convergence. Importantly, unlike the GBDA method, STP does not require introducing a reference model to guarantee concealment, ensuring its applicability across various scenarios.

On the other hand, the STP method is an optimization of the GCG method, with a notable distinction. Unlike the GCG method, STP considers the concealment of adversarial text. Given that certain online models employ input detection mechanisms to prevent malicious usage~\cite{deng2023jailbreaker}, traditional methods such as adding prefixes or suffixes in prompts are susceptible to perplexity detection. STP, in contrast, provides a less detectable avenue for jailbreak attacks, making it challenging to be identified.

\subsection{Truncation Protection Example}
\subsubsection{Different Models}
The specified end token is not the same for different models due to tokenizer differences. For example, the LLaMA model developed by Meta uses \textless{}/s\textgreater{} as the end token, which corresponds to a dictionary index of 2, while the end token in the cl100k-base~\cite{tiktoken} tokenizer of the GPT-4 model corresponds to a dictionary index of 100257. Therefore, different indexes should be used as the optimization target when constructing TPEs for different models.

For the work presented in this paper, achieving universality in truncation protection examples across models with different tokenizers proves challenging. Addressing this issue will be a key focus for future research efforts.

\subsubsection{Discussion on the Robustness of TPE}
\label{sec:Robustness}
In this section, we will discuss the robustness of TPE. Specifically, we selected the experimental results from Section \ref{sec:white-box}, concerning 120-token length texts from the Novel dataset optimized on the Vicuna model and subjected them to random word deletion, random word addition, and random synonym replacement operations~\cite{wei-zou-2019-eda}. The randomization ratio ranged from 5\% to 20\%. The experimental results are presented in Table \ref{tab:robust}, where $\text{RI}$ denotes random word insertion, $\text{RD}$ denotes random word deletion, and $\text{SR}$ denotes random synonym replacement, with subscripts indicating the random operation ratio.

\begin{table}[h]

\caption{Robustness Performance of TPE}
\label{tab:robust}
\centering
\begin{tabular}{|c|c|c|c|c|c|}
\hline
 \diagbox[innerwidth=2cm]{Metric}{Operations} & None & $\text{RI}_{5\%}$ & $\text{RI}_{10\%}$ & $\text{RI}_{15\%}$ & $\text{RI}_{20\%}$ \\ \hline 
 PSR & 0.80 & 0.51 & 0.39 & 0.33 & 0.27 \\ \hline \hline 
 \diagbox[innerwidth=2cm]{Metric}{Operations} & None & $\text{RD}_{5\%}$ & $\text{RD}_{10\%}$ & $\text{RD}_{15\%}$ & $\text{RD}_{20\%}$  \\ \hline 
 PSR & 0.80 & 0.48 & 0.33 & 0.23 & 0.15\\ \hline \hline
 \diagbox[innerwidth=2cm]{Metric}{Operations} & None & $\text{SR}_{5\%}$ & $\text{SR}_{10\%}$ & $\text{SR}_{15\%}$ & $\text{SR}_{20\%}$  \\ \hline 
 PSR & 0.80 & 0.51 & 0.37 & 0.29 & 0.25\\ \hline

\end{tabular}
\end{table}

The results indicate that the effectiveness of TPE protection deteriorates as the perturbation ratio increases. However, excessively high ratios of perturbation can cause significant information loss to the original text.

\subsubsection{Heightened Security Requirements}
For text with clearly defined security requirements, the TPE construction can be optimized by focusing on the prefixes and suffixes related to the specific topic. A new loss function can be constructed by introducing coefficients, and optimization should be applied only to the original text portion. For instance, when dealing with personal social media content, a defender may wish to conceal specific aspects of their lifestyle, and he can design corresponding prefixes and suffixes to construct TPE. The specific details are outlined in Algorithm~\ref{alg:Truncation-high}.

In this section, we prepared five rewrite-related prefixes, denoted as $\text{Prefix}_1'$, $\text{Prefix}_2'$, $\text{Prefix}_3'$, $\text{Prefix}_4'$, and $\text{Prefix}_5'$, along with 100 rewrite-related prefixes generated by ChatGPT-3.5. We defined the loss function as $(\sum_{i=1}^{5} \text{loss}_{\text{Prefix}_i'})/5$. The experimental parameters were set as $|S_i|=5$, $|N_i|=10$, $batch\ size=128$, and $T=15$. The experiments were conducted on the Vicuna model using texts with a length of 40 tokens from the Novel dataset. The experimental results are shown in Table~\ref{tab:instruction_transfer}, where $\text{PSR}^\dag$ represents the average results after transferring to the 100 rewrite-related prefixes. 

The experimental results show that under the theme of rewriting, TPE maintains a high PSR value even after transferring to unoptimized prefixes, with an average PSR value of 0.81 after transfer.

The specific prefixes are provided in the Appendix.

\begin{table}[h]
\caption{Transferability of Prefix Optimization}
\label{tab:instruction_transfer}
\centering
\begin{tabular}{@{}c|ccccc@{}}
\toprule
 \diagbox[innerwidth=1.5cm, outerrightsep=0cm, outerleftsep=0.1cm]{Metrics}{Prefix} & $\text{Prefix}_1'$ & $\text{Prefix}_2'$ & $\text{Prefix}_3'$ & $\text{Prefix}_4'$ & $\text{Prefix}_5'$ \\ \midrule 
  PSR & 0.88 & 0.91 & 0.89 & 0.90 & 0.91\\ \midrule 
 $\text{PSR}^\dag$ & \multicolumn{5}{c}{0.81}\\ \bottomrule
\end{tabular}
\end{table}

\subsubsection{Discussion on TPE Text Quality}
In the aforementioned STP method, we used n-nearest neighbors of tokens in the embedding layer to construct semantically similar candidate sets. In fact, the construction of semantically similar candidate sets in STP is a replaceable module, and different methods can be used to construct semantically similar candidate sets to improve the quality of TPE texts. In this section, we mainly discuss the impact of using BERT to construct~\cite{li2020bert} these candidate sets on the quality of TPE texts.

Specifically, we used the BERT-base-uncased model to construct semantically similar sets. Compared to the original method, this method improved the semantic fluency of the text without compromising optimization efficiency, although there remains a large semantic gap compared to the original text. Some experimental results on the Vicuna model are presented in Table \ref{tab:TPE-example-bert}.

\begin{table}[h!]
\caption{ Examples of TPE on Vicuna with BERT}
\label{tab:TPE-example-bert}
\centering
\begin{tabular}{@{}p{4cm}p{4cm}@{}}
\toprule
 \makecell[c]{Original Text} & \makecell[c]{Protected Text}\\ \midrule
 Can you \textcolor{blue}{help} me write a resignation letter to my \textcolor{blue}{current} employe\textcolor{blue}{r}, \textcolor{blue}{while} leaving on \textcolor{blue}{good} terms and expressing gratitude \textcolor{blue}{for} the opportunities provided? &
 Can you \textcolor{red}{exc} me \textcolor{red}{,} a resignation letter to my \textcolor{red}{previous} employ e\textcolor{red}{s}, \textcolor{red}{for} leaving on \textcolor{red}{our} terms and expressing gratitude \textcolor{red}{to} the opportunities provided? \\ \midrule

 Draft an apology email \textcolor{blue}{to} a \textcolor{blue}{customer} \textcolor{blue}{who} experienced a \textcolor{red}{delay} in their order, and \textcolor{blue}{provide} \textcolor{blue}{re}assurance that the issue \textcolor{blue}{has been resolved.} &
 Draft an apology email \textcolor{red}{by} a \textcolor{red}{student} \textcolor{red}{which} experienced a \textcolor{red}{change} in their order, and \textcolor{red}{seek} \textcolor{red}{an}assurance that his issue \textcolor{red}{previously remained cleared !} \\ \midrule
 
 Write a \textcolor{blue}{compe}lling product launch announ\textcolor{blue}{cement} email to inform \textcolor{blue}{our} \textcolor{blue}{customers} of our \textcolor{blue}{new} \textcolor{blue}{soft}ware \textcolor{blue}{solution.} &
 Write a \textcolor{red}{je}lling product launch announ \textcolor{red}{cy} email to inform \textcolor{red}{my} \textcolor{red}{staff} of our \textcolor{red}{current} \textcolor{red}{hard}ware \textcolor{red}{deployment ?} \\ \midrule
 
 Write a captivating \textcolor{blue}{movie} review for a \textcolor{blue}{recent}ly released \textcolor{blue}{science fiction} film, \textcolor{blue}{discuss}ing \textcolor{blue}{its plot}, characters, \textcolor{blue}{and special effects}. &
 Write a captivating \textcolor{red}{feature} review for a \textcolor{red}{new}ly released \textcolor{red}{action comedy} film, \textcolor{red}{learn}ing \textcolor{red}{about unusual situations}, characters \textcolor{red}{background \& software engineering}. \\ \bottomrule
\end{tabular}
\end{table}

\subsubsection{Universal Model}
For models that share the same tokenizer but have different parameters, the approach aligns with the universal method outlined in the GCG method\cite{zou2023universal}. This method enables simultaneous optimization for multiple models, thereby achieving enhanced TPE generality. The specific details are similar to Algorithm~\ref{alg:Truncation-high}, except that the loss function calculation involves the sum across different models.

\subsubsection{Other Truncation Method}
In addition to directly generating an end token as the first token, another strategy involves guiding the model to produce refusal responses. In previous work~\cite{zou2023universal}, ``Sure, here is" was used as an optimization prefix to guide the model during evasion attacks. Similarly, in this work, we can use refusal expressions like ``I am sorry, but" as optimization targets.

\subsection{Evasion Attack}
In this section, we will discuss the impact of three potential evasion attacks on TPE.
\subsubsection{Adversarial Prompt Engineering}
Against the TPE, an attacker can construct adversarial prompts to make the LLM regain the ability to output text. For example, an attacker can add text before and after the TPE to disrupt the TPE performance.

Effective adversarial prompts can be divided into two categories, one usually requiring more complex manual construction~\cite{liu2023jailbreaking}, while the other uses information such as gradients to optimize the construction process~\cite{shin2020autoprompt,zou2023universal,wallace2019universal,wen2023hard,guo2021gradient}.

For the first type of attack, the involvement of human labor makes the cost of the attack rise. If only pre-designed text is used to reach the automation goal, the attack will be less effective. The experiment in Section~\ref{sec:white-box} is similar to this scenario, which shows that the effect of adding prompts before and after the text on the performance of TPE decreases as the length of the TPE increases, and in addition, the addition of extraneous prompts may affect the model performance.

For the second type of attack, the attacker is able to construct adversarial prefixes and suffixes through gradient information, which in turn reduces the probability of sampling the end token in the first round. However, since lowering the probability of the end token does not correspond to the actual semantics, this approach may cause the LLM responses to deviate from the original question. In addition, this type of approach usually requires a large time overhead.
\subsubsection{Simple Text Perturbation}
Against the TPE, an attacker is able to degrade the protective effect of TPE through simple text perturbations. Examples include randomly adding words to the text, randomly deleting words from the text, randomly replacing words in the text, or segmenting the text.
The experiment in Section~\ref{sec:Robustness} is similar to this scenario, and the results show that increasing the text perturbation ratio can degrade the effectiveness of TPE protection, but it also causes information loss of original text, such as structural information, semantic information of the text.

\subsubsection{Retokenization}
Retokenization refers to encoding the inputs to the model in different tokenization methods~\cite{jain2023baseline}. In the simplest case, we break tokens apart and represent them using multiple smaller tokens. For example, the token ``studying" has a broken-token representation ``study"+``ing", among other possibilities.
Through the method of retokenization, an attacker is able to reduce the adversarial nature of the TPE, thus achieving the goal of reducing its performance. However, retokenization will degrade the performance of the model to some extent.
\begin{algorithm}[h]
    \caption{Truncation Protection Example for Heightened Security Requirements}
    \label{alg:Truncation-high}
    \DontPrintSemicolon
    \KwIn{Original Prompt $\mathcal{P}$, Iterations $T$, Target Model $M$, Batch Size $B$, Loss Function $\mathcal{L}$, Prefixes $pre_1,...,pre_m$, Suffixes $suf_1,...,suf_n$}
    \KwOut{Optimized prompt $\mathcal{P}$}
    \SetKwProg{Fn}{repeat}{}{}
    \Fn{T times}{
        \For{$i=1,...,m$}{
            $\mathcal{P}^i=pre_i+\mathcal{P}$\;
        }
        \For{$i=1,...,n$}{
            $\mathcal{P}^{i+m}=\mathcal{P}+suf_i$\;
        }
        $\text{loss}=0$\;
        \For{$i=1,...,m+n$}{
            $\text{loss} + = \eta_i \cdot \mathcal{L}(\mathcal{P}^i)$\;
        }
        \For{$i=1,...,len(\mathcal{P})$}{
            $h_i = -\nabla_{v_i} \text{loss}$\;
            $N_i=N(\mathcal{P}_i)$\;
            $S_i=Top-k(N_i)$\;
        }
        $\text{len}_{\text{part}}=\frac{B}{len(\mathcal{P})}$\;
        \For{$b=1,...,B$}{
            \uIf{$B > \text{len}(\mathcal{P})$}{$ i=\lceil\frac{b}{\text{len}_{\text{part}}}\rceil$}
        \uElse{
            \Repeat{$i \neq \text{previous } i$}{
                    $i = \text{random}(1, len(\mathcal{P}))$\;
                }
        }
            $\tilde{\mathcal{P}}^{(b)}=\mathcal{P}$\;
            $\tilde{\mathcal{P}}_i^{(b)}=Uniform(S_i)$\;
        }
        $\mathcal{P}=\tilde{\mathcal{P}}^{(b^*)}, \text{ where } b^* = \arg\min_b L(\tilde{\mathcal{P}}^{(b)})$\;
    }
    \Return $\mathcal{P}$ \;
\end{algorithm}

\section{Conclusion}
\label{sec:conclusion}
In this paper, we introduce the first text protection mechanism, SG, and the concept of TPE to address the problem of malicious exploitation of texts by LLMs. Meanwhile, we introduce the first method for constructing TPE, which is called STP. Our experimental outcomes demonstrated the effectiveness and concealment of the STP method across varied text lengths, types, and diverse models. Furthermore, TPE constructed using STP showed some level of transferability and robustness. Additionally, we explore the time overhead of constructing TPE using the STP method, which, under appropriately chosen parameters, incurs very short construction times for individual texts and exhibits a nearly linear relationship with text length. Additionally, we delved into optimizing specific token-associated model properties, which we believe can inspire future investigations into LLM characteristics. We aim to deploy SG in real-world scenarios to address the escalating concerns regarding LLM security and believe that it will find broader applications in the future.

\bibliography{bib.bib}{}
\bibliographystyle{IEEEtran}
\appendix
\begin{table}[h]
\renewcommand{\arraystretch}{0.2}
\caption{Rewrite Relevant Prefixes}
\centering
\scriptsize
\begin{tabular}{@{}p{8cm}@{}}
\toprule
 \makecell[c]{\textbf{Collected Prefixes}} \\ \midrule
 $\text{Prefix}_1'$: Please rewrite the following text. \\ \midrule
 $\text{Prefix}_2'$: Could you rephrase the following text? \\ \midrule
 $\text{Prefix}_3'$: I would appreciate it if you could restate the following text. \\ \midrule
$\text{Prefix}_4'$: Kindly reformulate the following text in a different way. \\ \midrule
$\text{Prefix}_5'$: Would you mind rewording the following text?
 \\ \midrule \midrule
\makecell[c]{\textbf{Other Rewrite-Related Prefixes}} \\ \midrule
 1. Could you please express the following text in a different way? \\ \midrule
2. I need you to paraphrase the following text. \\ \midrule
3. Please provide a different wording for the following text. \\ \midrule
4. Can you articulate the following text in another manner? \\ \midrule
5. Would you mind recasting the following text? \\ \midrule
6. Please convey the meaning of the following text in a different manner. \\ \midrule
7. Could you put the following text into other words? \\ \midrule
8. I require a rewording of the following text. \\ \midrule
9. Please restate the following text using different phrasing. \\ \midrule
10. Can you rephrase the following text in an alternative way? \\ \midrule
11. Kindly reformulate the following text in a different manner. \\ \midrule
12. Would you be able to reword the following text? \\ \midrule
13. Please provide an alternative expression for the following text. \\ \midrule
14. Could you articulate the following text in a different way? \\ \midrule
15. I need you to restate the following text. \\ \midrule
16. Please convey the essence of the following text in a different way. \\ \midrule
17. Can you express the following text in another manner? \\ \midrule
18. Would you mind rephrasing the following text? \\ \midrule
19. Please communicate the meaning of the following text in a different manner. \\ \midrule
20. Could you put the following text into different words? \\ \midrule
21. I require a rephrasing of the following text. \\ \midrule
22. Please restate the following text using alternative phrasing. \\ \midrule
23. Can you reword the following text in an alternate way? \\ \midrule
24. Kindly reformulate the following text in a different style. \\ \midrule
25. Would you be able to recast the following text? \\ \midrule
26. Please provide an alternate expression for the following text. \\ \midrule
27. Could you articulate the following text in a different manner? \\ \midrule
28. I need you to rewrite the following text. \\ \midrule
29. Please convey the substance of the following text in a different way. \\ \midrule
30. Can you express the following text in another form? \\ \midrule
31. Would you mind rephrasing the following text? \\ \midrule
32. Please communicate the essence of the following text in a different manner. \\ \midrule
33. Could you put the following text into various words? \\ \midrule
34. I require a rewording of the following text. \\ \midrule
35. Please restate the following text using diverse phrasing. \\ \midrule
36. Can you reword the following text in an alternative manner? \\ \midrule
37. Kindly reformulate the following text in a different fashion. \\ \midrule
38. Would you be able to recast the following text? \\ \midrule
39. Please provide an alternative articulation for the following text. \\ \midrule
40. Could you articulate the following text in a different way? \\ \midrule
41. I need you to rephrase the following text. \\ \midrule
\end{tabular}
\end{table}

\begin{table}[h]
\renewcommand{\arraystretch}{0.2}
\centering
\scriptsize
\begin{tabular}{@{}p{8cm}@{}}
\midrule
42. Please convey the content of the following text in a different manner. \\ \midrule
43. Can you express the following text in another style? \\ \midrule
44. Would you mind rewording the following text? \\ \midrule
45. Please communicate the substance of the following text in a different way. \\ \midrule
46. Could you put the following text into distinct words? \\ \midrule
47. I require a rephrasing of the following text. \\ \midrule
48. Please restate the following text using varied phrasing. \\ \midrule
49. Can you reword the following text in an alternate manner? \\ \midrule
50. Kindly reformulate the following text in a different manner. \\ \midrule
51. Would you be able to recast the following text? \\ \midrule
52. Please provide an alternate expression for the following text. \\ \midrule
53. Could you articulate the following text in a different manner? \\ \midrule
54. I need you to rewrite the following text. \\ \midrule
55. Please convey the essence of the following text in a different manner. \\ \midrule
56. Can you express the following text in another way? \\ \midrule
57. Would you mind rephrasing the following text? \\ \midrule
58. Please communicate the meaning of the following text in a different way. \\ \midrule
59. Could you put the following text into different words? \\ \midrule
60. I require a rewording of the following text. \\ \midrule
61. Please restate the following text using alternative phrasing. \\ \midrule
62. Can you reword the following text in an alternate way? \\ \midrule
63. Kindly reformulate the following text in a different style. \\ \midrule
64. Would you be able to recast the following text? \\ \midrule
65. Please provide an alternate articulation for the following text. \\ \midrule
66. Could you articulate the following text in a different fashion? \\ \midrule
67. I need you to rephrase the following text. \\ \midrule
68. Please convey the substance of the following text in a different manner. \\ \midrule
69. Can you express the following text in another form? \\ \midrule
70. Would you mind rewording the following text? \\ \midrule
71. Please communicate the essence of the following text in a different way. \\ \midrule
72. Could you put the following text into various words? \\ \midrule
73. I require a rephrasing of the following text. \\ \midrule
74. Please restate the following text using diverse phrasing. \\ \midrule
75. Can you reword the following text in an alternative manner? \\ \midrule
76. Kindly reformulate the following text in a different fashion. \\ \midrule
77. Would you be able to recast the following text? \\ \midrule
78. Please provide an alternative expression for the following text. \\ \midrule
79. Could you articulate the following text in a different way? \\ \midrule
80. I need you to rewrite the following text. \\ \midrule
81. Please convey the content of the following text in a different way. \\ \midrule
82. Can you express the following text in another style? \\ \midrule
83. Would you mind rephrasing the following text? \\ \midrule
84. Please communicate the substance of the following text in a different manner. \\ \midrule
85. Could you put the following text into distinct words? \\ \midrule
86. I require a rewording of the following text. \\ \midrule
87. Please restate the following text using varied phrasing. \\ \midrule
88. Can you reword the following text in an alternate manner? \\ \midrule
89. Kindly reformulate the following text in a different manner. \\ \midrule
90. Would you be able to recast the following text? \\ \midrule
91. Please provide an alternate expression for the following text. \\ \midrule
92. Could you articulate the following text in a different manner? \\ \midrule
93. I need you to rewrite the following text. \\ \midrule
94. Please convey the essence of the following text in a different manner. \\ \midrule
95. Can you express the following text in another way? \\ \midrule
96. Would you mind rephrasing the following text? \\ \midrule
97. Please communicate the meaning of the following text in a different way. \\ \midrule
98. Could you put the following text into different words? \\ \midrule
99. I require a rewording of the following text. \\ \midrule
100. Please restate the following text using alternative phrasing. \\ \bottomrule
\end{tabular}
\end{table}

\end{CJK}
\end{document}